\documentclass[letterpaper, 10 pt, conference]{ieeeconf}

\IEEEoverridecommandlockouts

\renewcommand{\baselinestretch}{0.945}

\usepackage{import}
\usepackage{wrapfig}
\usepackage[absolute,overlay]{textpos}

\usepackage{lipsum}
\usepackage{times}
\usepackage{bm}

\usepackage[usenames,dvipsnames]{xcolor}

\usepackage{amsmath,amssymb,mathtools}

\usepackage{amsthm}
\usepackage{bbold,dsfont,bbm}

\usepackage{graphicx}
\usepackage{tikz,tikz-cd,pgfplots}
\usepgfplotslibrary{fillbetween}
\usepackage{paralist}

\usepackage{hyperref}
\usepackage[capitalise,noabbrev]{cleveref}
\usepackage{cite}

\usepackage{multirow}
\usepackage{booktabs}

\usepackage[acronym]{glossaries}

\usepackage{comment}

\usepackage{units}

\usepackage[font=footnotesize]{caption}
\usepackage{subcaption}

\newacronym{abk:av}{AV}{autonomous vehicle}

\newacronym{abk:cpo}{CPO}{complete partial order}
\newacronym{abk:cdp}{CDP}{co-design problem}
\newacronym{abk:cdpi}{CDPI}{co-design problem with implementation}

\definecolor{baiocchi}{RGB}{193,221,245}
\newcommand*\circled[1]{\tikz[baseline=(char.base)]{
            \node[shape=circle,black,inner sep=1pt, fill=baiocchi] (char) {#1};}}

\definecolor{dpred}{rgb}{0.7, 0.0, 0.0}

\newcommand{\dis}[1]{{#1}_\mathrm{d}}

\newacronym{abk:dp}{DP}{design problem}
\newacronym{abk:dpi}{MDPI}{Monotone design problem with implementation}
\newacronym{abk:dcpo}{DCPO}{directed complete partial order}

\newcommand{\effort}{P_\mathrm{effort}}

\definecolor{dpgreen}{rgb}{0.0, 0.5, 0.0}
\newcommand{\F}[1]{{\color{dpgreen}#1}}
\newcommand{\Ftext}[1]{\textcolor{dpgreen}{#1}}

\newcommand{\mat}[1]{\mathbf{#1}}

\newcommand{\op}{^{\mathrm{op}}}

\def\prov{\mathsf{prov}}
\newacronym{abk:poset}{poset}{partially ordered set}
\newcommand{\posdef}{\mathcal{P}^+}
\newcommand{\possemidef}{\mathcal{P}}

\newcommand{\Rtext}[1]{\textcolor{dpred}{#1}}

\newcommand{\R}[1]{{\color{dpred}#1}}

\def\req{\mathsf{req}}

\newcommand{\setOfFunctionalities}[1]{\F{\mathcal{F}_{#1}}}
\newcommand{\setOfFunctionalitiesOp}[1]{\F{\mathcal{F}_{#1}}^{\mathrm{op}}}
\newcommand{\setOfImplementations}[1]{\mathcal{I}_{#1}}
\newcommand{\setOfResources}[1]{\R{\mathcal{R}_{#1}}}

\newcommand{\trace}[1]{\mathsf{Tr}\left(#1\right)}

\newcommand{\tup}[1]{\langle#1\rangle}

\newcommand{\track}{P_\mathrm{track}}

\newcommand{\D}{\mathrm{d}}

\newtheorem{theorem}{Theorem}
\newtheorem{lemma}{Lemma}
\theoremstyle{definition}

\newtheorem{definition}{Definition}
\newtheorem{example}{Example}
\theoremstyle{remark}
\newtheorem*{remark}{Remark}

\makeatletter
    \if@cref@capitalise
        \crefname{subsection}{Section}{Sections}
        \crefname{subsubsection}{Section}{Sections}
        \crefname{assump}{Assumption}{Assumptions}
        \crefname{problem}{Problem}{Problems}
    \else
        \crefname{subsection}{section}{sections}
        \crefname{subsubsection}{section}{sections}
        \crefname{assump}{assumption}{assumptions}
        \crefname{problem}{problem}{problems}
    \fi
\makeatother

\usetikzlibrary{positioning,intersections, hobby, patterns, calc, decorations.pathmorphing, decorations.markings, shadows,shapes, cd,arrows.meta, fit,quotes}

\tikzset{
   tick/.style={postaction={
      decorate,
      decoration={markings, mark=at position 0.5 with {\draw[-] (0,.4ex) -- (0,-.4ex);}}}
   }
}
\tikzstyle{block} = [draw, rectangle, minimum height=2em, minimum width=3em,font=\bfseries,rounded corners,thick]
\tikzstyle{block} = [draw, rectangle, minimum height=2em, minimum width=3em]
\tikzstyle{block1} = [draw, rectangle, minimum height=1.5em, minimum width=2.5em]
\tikzstyle{blockDyn} = [draw, rectangle, minimum height=2.5em, minimum width=3.5em, align=center, inner sep=10pt, thick, fill=white, copy shadow={draw=black,fill=black,opacity=1,shadow xshift=0.5ex,shadow yshift=-0.5ex}]
\tikzstyle{blockAlg} = [draw, rectangle, minimum height=1.5em, minimum width=2.5em, align=center, inner sep=10pt, thick]
\tikzstyle{sum} = [draw,circle]

\tikzstyle{nodePre} = [circle, draw,inner sep=1pt,node contents={$\preceq$},thick]
\tikzstyle{nodePreEmpty} = [circle, draw,inner sep=1pt,thick]
\tikzstyle{nodePos} = [circle, draw,inner sep=1pt,node contents={$\posceq$},thick]
\tikzstyle{nodeProd} = [rectangle, draw,inner sep=4pt,node contents={$\times$},rounded corners,thick]
\tikzstyle{nodeSum} = [rectangle, draw,inner sep=4pt,node contents={$\mathbf{+}$},rounded corners,thick]

\definecolor{red}{rgb}{0.75, 0.0, 0.0}

\tikzset{fcname/.store in =\fcname, fcname={}}
\tikzset{funame/.store in =\funame, funame={}}
\tikzset{rcname/.store in =\rcname, rcname={}}
\tikzset{runame/.store in =\runame, runame={}}
\tikzset{whereres/.store in =\whereres, whereres=0.5}
\tikzset{wherefun/.store in =\wherefun, wherefun=0.5}
\tikzset{relres/.store in =\relres, relres={above}}
\tikzset{relfun/.store in =\relfun, relfun={above}}
\tikzset{posres/.store in =\posres, posres=1}
\tikzset{posfun/.store in =\posfun, posfun=1}
\tikzset{loos/.store in =\loos, loos=2}
\tikzset{feedback/.store in =\feedback, feedback=0}
\tikzset{
   DP/.style={%
      label/.style={
         font=\everymath\expandafter{\the\everymath\scriptstyle},
         inner sep=5pt,
         node distance=2pt and -2pt},
      semithick,
      node distance=1 and 1,
      rconn/.style={color=white,opacity=0.0,postaction={decorate}, shorten <=3.2pt, shorten >= 0.8,
      decoration={markings, 
      mark= at position 0 with {
               \coordinate (a);
      },
      mark=at position .5 with
      {
              \ifthenelse{\equal{\feedback}{1}}{\def\angleOut{90}\def\angleIn{90}}{\def\angleOut{0}\def\angleIn{180}}    
              \coordinate (b);
              \draw[dashed,dpred,opacity=1.0] (a) to[out=\angleOut,in=\angleIn,looseness=\loos] 
              node[pos=\posres,\relres=\whereres mm,dpred,opacity=1,fill=white,inner sep=1pt,outer sep=1pt]{\footnotesize{\rcname}} (b);
      },
      mark= at position 1 with 
      {
             \ifthenelse{\equal{\feedback}{1}}{\def\angleOut{0}\def\angleIn{0}}{\def\angleOut{180}\def\angleIn{0}} 
              \ifthenelse{\equal{\feedback}{1}}{\def\symbol{\succeq}}{\def\symbol{\preceq}} 
              \coordinate (c);
              \draw[dpgreen,opacity=1.0] (c) to[out=\angleOut,in=\angleIn,looseness=\loos]
              node[pos=\posfun,\relfun=\wherefun mm,dpgreen,opacity=1,fill=white,inner sep=1pt,outer sep=1pt]{\footnotesize{\fcname}} (b){}; %
              \node[draw,circle,inner sep=0.5pt,color=black,fill=white,opacity=1.0] at (b) (nodepreceq) {$\symbol$}; 
      }
      }},
      runconn/.style={color=dpred,dashed,postaction={decorate},
      decoration={markings,
      mark= at position 1 with {
              \coordinate (a);
              \draw[dpred,opacity=1.0,dashed] ($(a)+(0.05,0)$) --++ (0.5,0) node[\relres,pos=\posres]{\footnotesize{\runame}};}
      }
      },
      funconn/.style={color=white,postaction={decorate},
      decoration={markings,
      mark= at position 0 with {
      \coordinate (a);
      \draw[dpgreen] ($(a)+(-0.05,0)$) -- ($(a)+(-0.5,0)$) node[\relfun, pos=\posfun]{\footnotesize{\funame}};}
      }
      },
      execute at begin picture={\tikzset{
         x=\dpx, y=\dpy,
         every fit/.style={inner xsep=\dpx, inner ysep=\dpy}}}
      },
   dpx/.store in=\dpx,
   dpx = 1.5cm,
   dpy/.store in=\dpy,
   dpy = 1.5ex,
   dp port sep/.store in=\dpportsep,
   dp port sep=2,
   dp port length/.store in=\dpportlen,
   dp port length=4pt,
   dp min width/.store in=\dpminwidth,
   dp min width=0.5cm,
   dp rounded corners/.store in=\dpcorners,
   dp rounded corners=2pt,
   dp small/.style={dp port sep=1, dp port length=2.5pt, dpx=.4cm, dp min width=.4cm, dpy=.7ex},
   dp/.code 2 args={%
      \pgfmathsetlengthmacro{\dpheight}{\dpportsep * (max(#1,#2)) * \dpy}
      \pgfkeysalso{draw,%
        minimum width=\dpminwidth,%
        minimum height=\dpheight,%
        font=\bfseries,
        outer sep=0pt,%
        inner sep=5pt,%
        rounded corners=\dpcorners,
        thick,
        prefix after command={\pgfextra{\let\fixname\tikzlastnode}},
        append after command={\pgfextra{\draw
            \ifnum #1=0{} \else foreach \i in {1,...,#1} { 
            ($(\fixname.north west)!{\i/(#1+1)}!(\fixname.south west)$) +(0,0) node[solid,left,circle,color=dpgreen,draw,fill=dpgreen,scale=0.3] {} coordinate (\fixname_fun\i) -- +(0,0) coordinate (\fixname_fun\i')}\fi %
            \ifnum #2=0{} \else foreach \i in {1,...,#2} {
            ($(\fixname.north east)!{\i/(#2+1)}!(\fixname.south east)$) +(0,0) coordinate (\fixname_res\i') -- +(0,0) node[solid,right,circle,color=dpred,draw,fill=dpred,scale=0.3] {} coordinate (\fixname_res\i)}\fi;
         }}}
         },
      dp name/.style={append after command={\pgfextra{\node[label=center,inner sep=2pt,fill=white] at (\fixname) {\textbf{#1}};}}}
   }

\crefname{equation}{}{}
\crefname{figure}{Fig.}{}
\crefname{definition}{Def.}{}

\newboolean{extended}
\setboolean{extended}{true}
\ifthenelse{\boolean{extended}}{
\title{
\textbf{Co-Design of Autonomous Systems: From Hardware Selection to Control Synthesis}
}
\author{Gioele Zardini, Andrea Censi, Emilio Frazzoli
\thanks{
This work was supported by the Swiss National Science Foundation under NCCR Automation, grant agreement 51NF40\_180545. The authors are with the Institute for Dynamic Systems and Control, ETH Z\"urich, Switzerland.
{\tt \{gzardini,acensi,efrazzoli\}@ethz.ch}.}
}}
{
\title{
\textbf{Co-Design of Autonomous Systems: From Hardware Selection to Control Synthesis}
}
\author{Gioele Zardini, Andrea Censi, Emilio Frazzoli
\thanks{
This work was supported by the Swiss National Science Foundation under NCCR Automation, grant agreement 51NF40_180545.
The authors are with the Institute for Dynamic Systems and Control, ETH Z\"urich, Switzerland.
{\tt \{gzardini,acensi,efrazzoli\}@ethz.ch}.}
}}

\begin{document}

\begin{textblock*}{\textwidth}(15mm,18mm) %
\bf \textcolor{NavyBlue}{To appear in the Proceedings of the 20th European Control Conference (ECC), 2021.}
\end{textblock*}

\maketitle
\begin{abstract}
Designing cyber-physical systems is a complex task which requires insights at multiple abstraction levels. The choices of single components are deeply interconnected and need to be jointly studied. In this work, we consider the problem of co-designing the control algorithm as well as the platform around it. In particular, we leverage a monotone theory of co-design to formalize variations of the LQG control problem as monotone feasibility relations. We then show how this enables the embedding of control co-design problems in the higher level co-design problem of a robotic platform. We illustrate the properties of our formalization by analyzing the co-design of an autonomous drone performing search-and-rescue tasks and show how, given a set of desired robot behaviors, we can compute Pareto efficient design solutions. 
\end{abstract}

\ifthenelse{\boolean{extended}}{
\section{Introduction}
\label{sec:introduction}

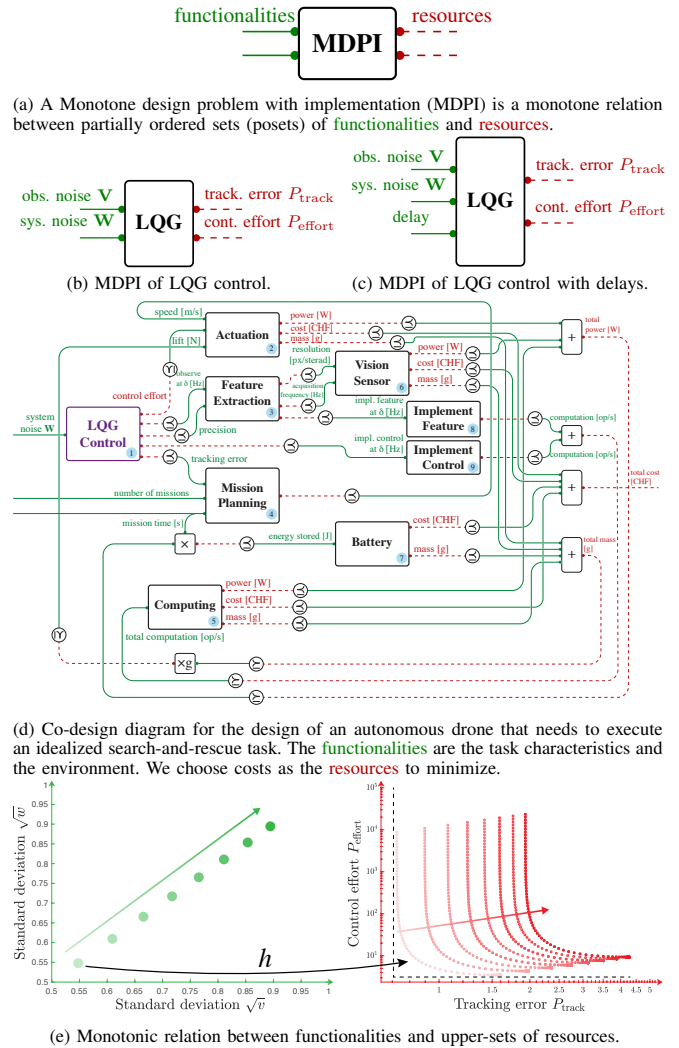
\begin{figure}[h!]
    \begin{subfigure}{\columnwidth}
    \begin{center}
    \begin{tikzpicture}[DP]
            \node[dp={2}{2}] (cnt) {MDPI};
            \draw[runconn, runame={resources}, relres=above,posres=0.9] (cnt_res1){};
            \draw[runconn, runame={}, relres=above,posres=0.9] (cnt_res2){};
            \draw[funconn, funame={functionalities},relfun=above,posfun=1.15] (cnt_fun1){};
            \draw[funconn, funame={},relfun=above,posfun=1.15] (cnt_fun2){};
\end{tikzpicture}
    \subcaption{A \gls{abk:dpi} is a monotone relation between partially ordered sets (posets) of \Ftext{functionalities} and \Rtext{resources}. \label{fig:mathcodesign}}
    \end{center}
    \end{subfigure}
    ~
    \begin{subfigure}{0.49\columnwidth}
    \begin{center}
    \scalebox{0.8}{\begin{tikzpicture}[DP, dp port sep =3]
            \node(ref) at (0,0){};
            \node[dp={2}{2},below=0.5cm of ref] (cnt) {LQG};
            \draw[runconn, runame={track. error $\track$}, relres=above, posres=1.6] (cnt_res1){};
            \draw[runconn, runame={cont. effort $\effort$},relres=above,posres=1.6] (cnt_res2){};
            \draw[funconn, funame={obs. noise $\mat{V}$},relfun=above, posfun=1.3] (cnt_fun1){};
            \draw[funconn, funame={sys. noise $\mat{W}$},relfun=above, posfun=1.3] (cnt_fun2){};
\end{tikzpicture}}
    \vspace{-0.5cm}
    \subcaption{\gls{abk:dpi} of LQG control. \label{fig:dp_ct}}
    \end{center}
    \end{subfigure}
    \begin{subfigure}{0.49\columnwidth}
    \begin{center}
    \scalebox{0.8}{\begin{tikzpicture}[DP, dp port sep =3]
            \node[dp={3}{2}] (cnt) {LQG};
            \draw[runconn, runame={track. error $\track$}, relres=above, posres=1.6] (cnt_res1){};
            \draw[runconn, runame={cont. effort $\effort$},relres=above,posres=1.6] (cnt_res2){};
            \draw[funconn, funame={obs. noise $\mat{V}$},relfun=above, posfun=1.3] (cnt_fun1){};
            \draw[funconn, funame={sys. noise $\mat{W}$},relfun=above, posfun=1.3] (cnt_fun2){};
            \draw[funconn, funame={delay},relfun=above, posfun=1] (cnt_fun3){};
\end{tikzpicture}}
    \vspace{-0.5cm}
    \subcaption{\gls{abk:dpi} of LQG control with delays. \label{fig:dp_ct_del}}
    \end{center}
    \end{subfigure}
    \begin{subfigure}{\columnwidth}
    \begin{center}
    \includegraphics[width=\columnwidth]{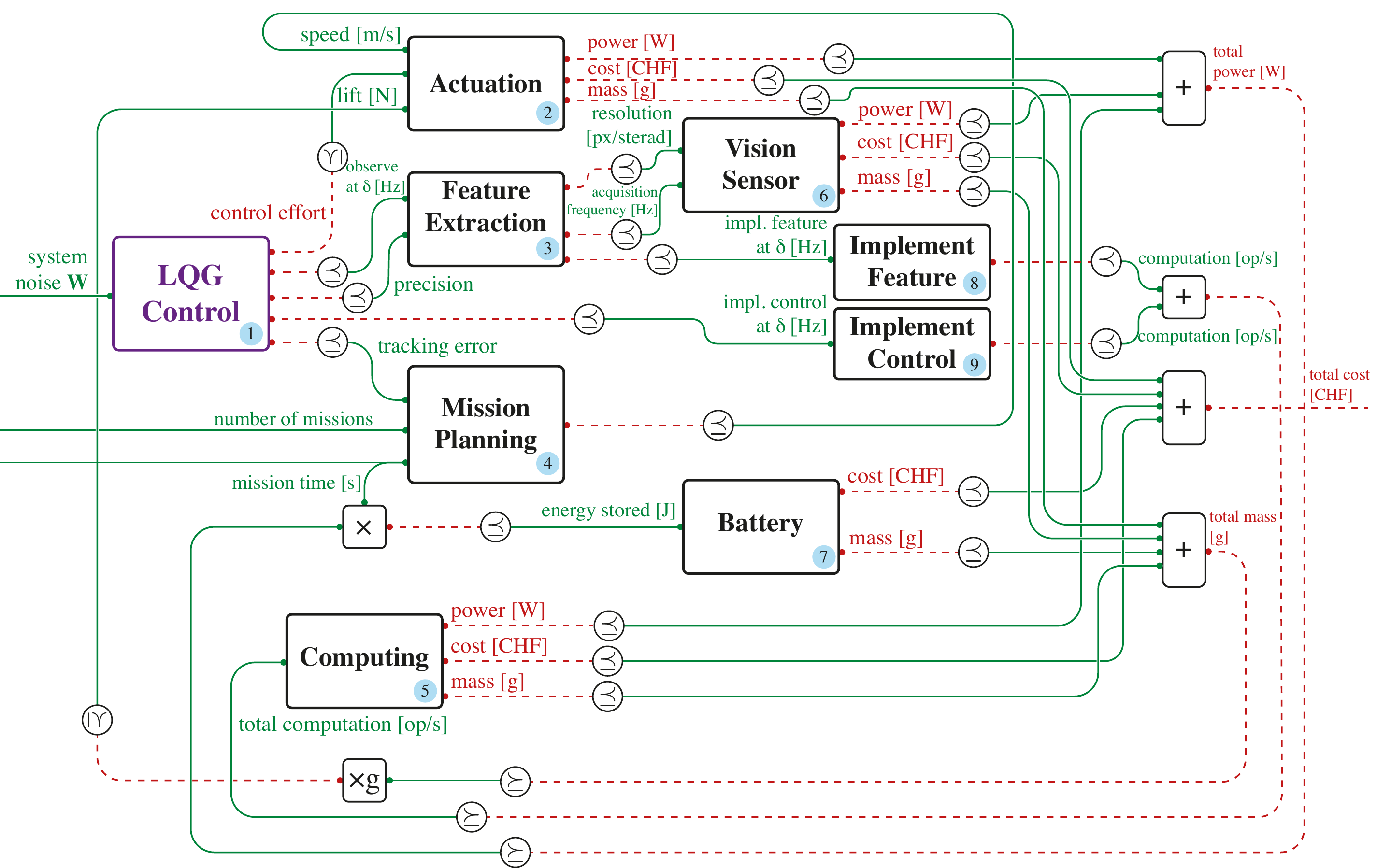}
    \subcaption{Co-design diagram for the design of an autonomous drone that needs to execute an idealized search-and-rescue task. The \Ftext{functionalities} are the task characteristics and the environment. We choose costs as the \Rtext{resources} to minimize. \label{fig:drone_codesign}}
    \end{center}
    \end{subfigure}
    \begin{subfigure}[b]{\columnwidth}
    \begin{center}
    \includegraphics[width=\columnwidth]{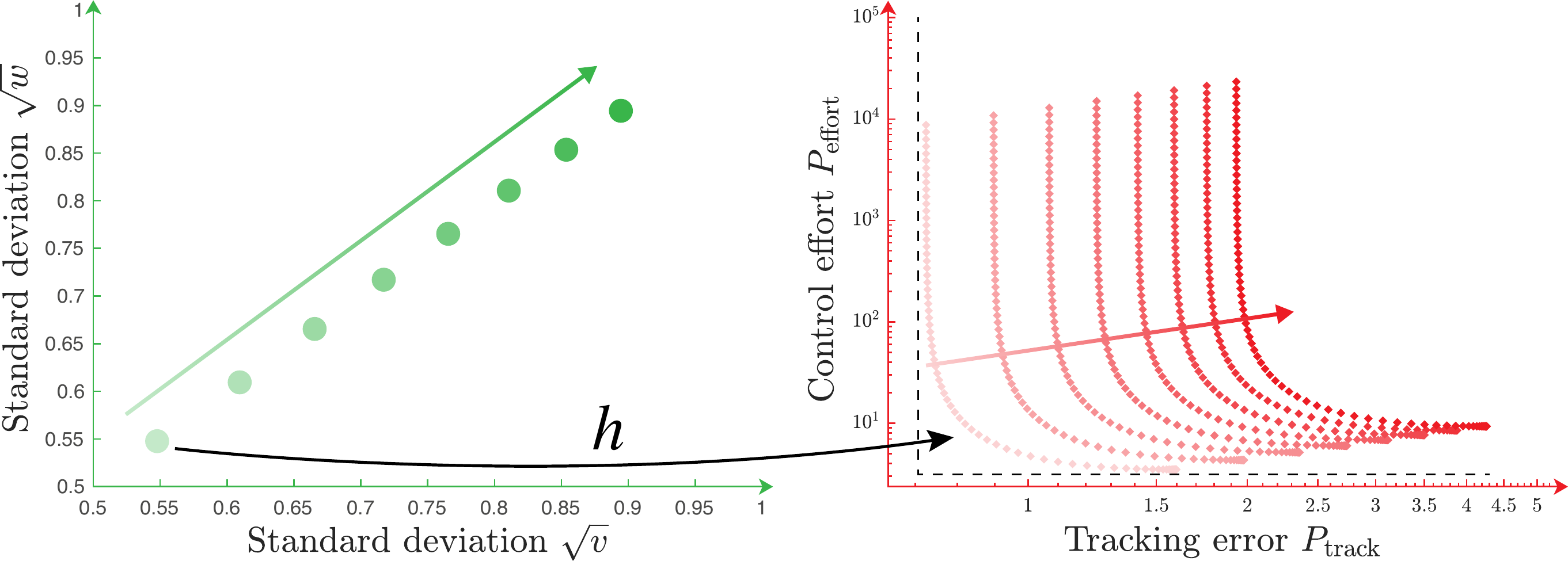}
    \subcaption{Monotonic relation between functionalities and upper-sets of resources. \label{fig:pareto}}
    \end{center}
    \end{subfigure}
    \caption{In this work we show how to use a monotone theory of co-design to embed LQG control design in a \gls{abk:dpi} for an entire robotic platform. (a-c): We show how to formalize (variations of) LQG control in the co-design framework, highlighting \Ftext{functionalities}, \Rtext{resources} and monotonic relations between them. (d): We then show how to embed the LQG \gls{abk:dpi} in a higher-level co-design diagram. Compositionality assures that if all blocks are \glspl{abk:dpi}, their composition is a \gls{abk:dpi}. (e): Co-design optimization techniques allow to obtain \Rtext{resource}-\Ftext{functionality} trade-offs in form of Pareto fronts.}
    \label{fig:firstsight}
    \vspace{-6mm}
\end{figure}

\begin{figure}[t]
\begin{subfigure}[b]{\columnwidth}
\begin{center}
\includegraphics[width=0.65\columnwidth]{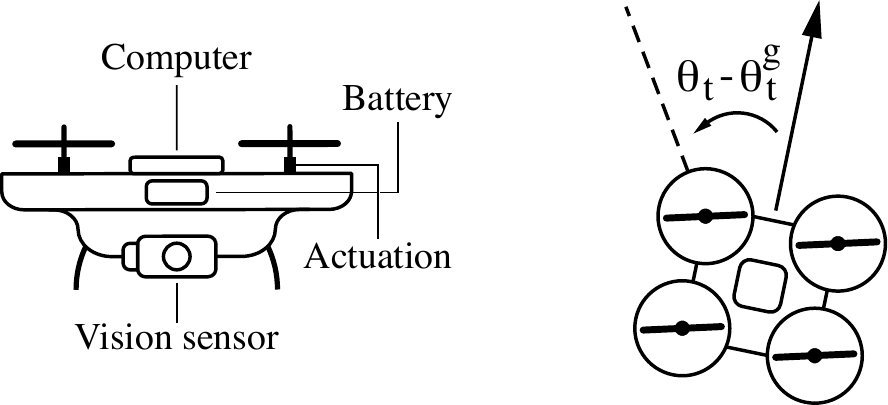}
\end{center}
\end{subfigure}
\caption{We consider the case study of a drone which needs to perform search-and-rescue tasks, and control its alignment with a given goal.}
\label{fig:drone_pics}
\vspace{-6mm}
\end{figure}

Cyber-physical systems are composed of material parts, such as sensors, actuators and computing units, and software components, such as perception, planning, and control modules. Traditionally, the design optimization of such components is treated in a compartmentalized manner, assuming specific, fixed configurations for the parts of the system which are not object of the design exercise. However, the designs of both physical and software components are deeply interconnected and need to be jointly considered. Examples of the importance of a joint design of components range from power-constrained robot design~\cite{censi2015power,ichnowski2020} to the design of fleets of autonomous vehicles which need to provide mobility services jointly with the existing urban transportation system~\cite{spieser2014toward, ZardiniLanzettiEtAl2020b}. Open questions are: What is the simplest sensor which a robot can use to guarantee a specific performance in a given task? How will this choice affect computation? Which platform design minimizes the overall monetary cost?

In this work, we propose an approach for the co-design of the platform and its control system~(\cref{fig:drone_codesign}), with particular reference to robotics. While a broad theory to optimize the control of specific systems has been developed, there exists no formalism which allows to jointly design control strategies and the platforms on which they will be executed. This is an issue, since robotic platforms are typically not fixed, and components such as actuators, sensors and batteries are subject to continuous changes.

\paragraph*{Related literature} Traditionally, researchers have been working on specific instances of the co-design of control, sensing and actuation~\cite{merlet2005,okane2008,maasoumy2013co,soudbakhsh2013co,zhu2018codesign,zheng2016cross,joshi2008,gupta2006,tzoumas2020,zhang2020,tatikonda2004,tanaka2015,karaman2012,moreno2015,guo2019,ballotta2020computation,zardiniICRA2020, bakirtzis2020categorical}.
The author of~\cite{merlet2005} formalizes the optimal design of the robotic platform for serial manipulators, and a precise analysis of trade-offs in robot sensing and actuation for worst-case scenarios is provided in~\cite{okane2008}. The authors of~\cite{maasoumy2013co} and \cite{soudbakhsh2013co} propose approaches for the co-design of control algorithms and platforms, applied to lane keeping and HVAC systems, respectively, and \cite{zhu2018codesign,zheng2016cross} focus on performance and security of cyber-physical systems. The problem of sensor selection typically cannot be solved in a closed form, but~\cite{joshi2008,gupta2006} show that in particular cases, there is enough structure for efficient optimization. Specifically, the authors of~\cite{tzoumas2020} study the problem of sensing-constrained LQG control, providing examples of robotic tasks, and authors of~\cite{zhang2020} optimally select sensors to solve path planning problems. In~\cite{tatikonda2004}, the authors propose an approach for optimal control with communication constraints and~\cite{tanaka2015} provides a framework to jointly optimize sensor selection and control, by minimizing the information acquired by the sensor. An analysis of performance limits for robotic tasks parametrized by the complexity of the environment is given in~\cite{karaman2012}. Selection of sensors and actuators is studied in~\cite{moreno2015}, where the authors optimize the design of the system with the goal of providing robust control for aeroservoelastic systems. Furthermore,~\cite{guo2019} proposes an approach to optimize actuators placement in large-scale networks control,~\cite{ballotta2020computation} studies computation-communication trade-offs and sensor selection for processing networks, and~\cite{bakirtzis2020categorical} gives a broad view of compositional cyber-physical system theories. 
In conclusion, to the best of our knowledge, existing design techniques for autonomous platforms are characterized by a fixed, problem-specific structure and do not allow to formalize and solve system co-design problems involving control synthesis.

\paragraph*{Statement of contributions}
In this work we show how to use a monotone theory of co-design to frame the design of LQG control strategies in the co-design problem of an entire robotic platform.
First, we formulate (variations of) LQG control as a \gls{abk:dpi}, characterized by feasibility relations between the control performance, accuracy, measurements' and system's noises. Second, we show how nuisances such as delays, digitalization and intermittent observations can be integrated in our formalism. The proofs for our results are available in~\cref{sec:app_A,sec:app_B}. Finally, we showcase the advantages of our approach, considering the co-design problem of a mobile robot which needs to execute an idealized search-and-rescue task (\cref{fig:drone_codesign}). We illustrate how, given the robot model and desired behaviors, we can formulate and answer several questions regarding the design of the robot.

\section{Monotone Co-Design Theory}
\label{sec:preliminaries}
We assume the reader to be familiar with basic concepts from order theory (listed in \cref{sec:app_order}). A possible reference is~\cite{davey2002}.
In this section, we report the main concepts related to the mathematical theory of co-design~\cite{censi2015,censi2016}. 

\begin{definition}[\gls{abk:dpi}]
\label{def:dpi}
Given the \glspl{abk:poset} $\setOfFunctionalities{},\setOfResources{}$, representing \Ftext{functionalities} and \Rtext{resources}, respectively, we define a  \emph{\gls{abk:dpi}} $d$ as a tuple $\tup{\setOfImplementations{d},\prov,\req}$, where $\setOfImplementations{d}$ is the \emph{set of implementations}, and $\prov$, $\req$ are functions mapping $\setOfImplementations{d}$ to $\setOfFunctionalities{}$ and $\setOfResources{}$, respectively:
\begin{equation*}
        \setOfFunctionalities{} \xleftarrow{\prov} \setOfImplementations{d} \xrightarrow{\req} \setOfResources{}.
\end{equation*}
To each \gls{abk:dpi} we associate a monotone map $\bar{d}$ given by
\begin{equation*}
    \begin{aligned}
        \bar{d}\colon \setOfFunctionalitiesOp{} \times \setOfResources{} &\to \mathbb{P}({\setOfImplementations{d}})\\
        \langle \F{f}^*,\R{r}\rangle &\mapsto \{i \in \setOfImplementations{d}\colon (\prov(i) \succeq_{\setOfFunctionalities{}}\F{f}) \wedge (\req(i)\preceq_{\setOfResources{}}\R{r})\},
    \end{aligned}
\end{equation*}
where~$(\cdot)\op$ reverses the order of a \gls{abk:poset}. A \gls{abk:dpi} is represented in diagrammatic form as in~\cref{fig:mathcodesign}. The expression~$\bar{d}(\F{f^*},\R{r})$ returns the set of implementations $S\subseteq \mathcal{I}_d$ for which \Ftext{functionalities} $\F{f}$ are feasible with \Rtext{resources} $\R{r}$. 
\end{definition}

\begin{example}[Monotonicity]
\label{ex:computing}
Consider the \gls{abk:dpi} of a \textbf{battery} (block \circled{7} in~\cref{fig:drone_codesign}) by means of the provided \Ftext{energy} and the required \Rtext{cost} and \Rtext{mass}. If a set $S=d(\F{f}^*,\R{r})$ of batteries require $\R{r}=\tup{\R{\text{mass}}, \R{\text{cost}}}$ to provide $\F{f}=\F{\text{energy}}$, then they can also provide less energy $\F{f'}\preceq \F{f}$, i.e. $S'\supseteq S$. On the other hand, if one has more resources $\R{r''}\succeq \R{r}$, the new set of batteries will at least still provide $\F{f}$, i.e. $S''\supseteq S$.
\end{example}

\begin{definition}
\label{def:map_h}
Given a \gls{abk:dpi} $d$, we define monotone maps $h_d\colon \setOfFunctionalities{}\to \mathcal{A} \setOfResources{}$, mapping a functionality to the minimum antichain of resources providing it, and $h_{d}'\colon \setOfResources{}\to \mathcal{A}\setOfFunctionalities{}$, mapping a resource to the maximum antichain of functionalities provided by it. To solve \glspl{abk:dpi} we find these maps, relying on Kleene's fixed point theorem~\cite{censi2015}. 
\end{definition}

Individual \glspl{abk:dpi} can be composed in several ways to form a co-design problem (\cref{fig:dpcompositions}). Series composition describes the case in which the functionality of a \gls{abk:dpi} becomes the resource of another \gls{abk:dpi}. For instance, the computation power provided by a computer is required by algorithm implementations. The relation ``$\preceq$'' appearing in \cref{fig:dpcompositions} represents a co-design constraint: The resource one component requires has to be at most the one provided by another component. Parallel composition corresponds to processes happening together. Finally, loop composition describes feedback.\footnote{It can be proved that the formalization of feedback makes the category of \glspl{abk:dpi} a traced monoidal category~\cite{fong2019}. 
}
Composition operations preserve monotonicity and thus all related algorithmic properties~\cite{censi2015}.
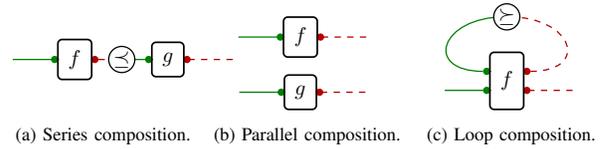
\begin{figure}[tb]
\begin{center}
\begin{subfigure}[b]{0.3\columnwidth}
\centering
\scalebox{0.8}{\begin{tikzpicture}[DP]
    \node[dp={1}{1}] (f) {$f$};
    \node[dp={1}{1}, right=1cm of f] (g) {$g$};
    \draw[rconn, rcname={}, fcname={}] (f_res1)  to (g_fun1);
    \draw[runconn, runame={}] (g_res1);
    \draw[funconn, funame={}] (f_fun1);
\end{tikzpicture}}
\subcaption{Series composition.}
\end{subfigure}
\begin{subfigure}[b]{0.3\columnwidth}
\centering
\scalebox{0.8}{\begin{tikzpicture}[DP]
    \node[dp={1}{1}] (f) {$f$};
    \node[dp={1}{1}, below=0.3cm of f] (g) {$g$};
    \draw[runconn, runame={}] (f_res1){};
    \draw[runconn, runame={}] (g_res1){};
    \draw[funconn, funame={}] (f_fun1){};
    \draw[funconn, funame={}] (g_fun1){};
\end{tikzpicture}}
\subcaption{Parallel composition.}
\end{subfigure}
\begin{subfigure}[b]{0.3\columnwidth}
\centering
\scalebox{0.8}{\begin{tikzpicture}[DP]
    \node[dp={2}{2}] (f) {$f$};
    \draw[runconn, runame={}] (f_res2){};
    \draw[funconn, funame={}] (f_fun2){};
    \draw[rconn,rcname={},fcname={},feedback=1,loos=3] (f_res1) -- ($(f)+(0,5)$) |- (f_fun1);
\end{tikzpicture}}
\subcaption{Loop composition.}
\end{subfigure}
\label{fig:diagrams}
\caption{\glspl{abk:dpi} can be composed in different ways.}
\label{fig:dpcompositions}
\end{center}
\vspace{-5mm}
\end{figure}
\section{Continuous-time LQG control as a \gls{abk:dpi}}
\label{sec:part_i}
\paragraph*{Background on LQG control}
\noindent First, we recall the definition of infinite-horizon LQG control.
\begin{definition}[Continuous-time LQG control]
\label{def:ctlqg}
Given the continuous-time stochastic dynamics
\begin{equation*}
\label{eq:cont_dyn}
    \begin{aligned}
    \D \bm{x}_t&=\mat{A}\bm{x}_t\D t+\mat{B}\bm{u}_t \D t+\mat{E} \D \bm{w}_t\\
    \D \bm{y}_t &= \mat{C}\bm{x}_t \D t+\mat{G}\D \bm{v}_t,
    \end{aligned}
\end{equation*}
where $\bm{v}_t$ and $\bm{w}_t$ are two standard Brownian processes, let $\mat{A},\mat{B},\mat{C}$, $\mat{E}$, $\mat{G}$ be matrices of compatible dimensions and $\mat{W}=\mat{E}\mat{E}^*$, $\mat{V}=\mat{G}\mat{G}^*$ be the effective noise covariances. The continuous-time infinite-horizon \emph{LQG problem} consists of finding a control law $\bm{u}^\star$ minimizing the quadratic cost
\begin{equation*}
\label{eq:general_continuous_cost}
    J=\textstyle{\lim_{T\to \infty}} \frac{1}{T}\mathbb{E}\{\smallint_{0}^{T} \left(\left( \bm{x}_t^\intercal \mat{Q} \bm{x}_t\right)+\left( \bm{u}_t^\intercal \mat{R}\bm{u}_t\right)\right) \D t\},
\end{equation*}
with $\mat{Q}\in \possemidef$, $\mat{R}\in \posdef$, where $\posdef$ ($\possemidef$) is the set of positive definite (positive semi-definite) matrices.
\end{definition}

\begin{remark}
In this work we use the Hermitian matrices \gls{abk:poset}~$\tup{\mathcal{M}^n,\preceq}$.
Given two Hermitian matrices of order~$n$ $\mat{A},\mat{B}\in \mathcal{M}^n$, we have $\mat{A}\preceq \mat{B} \Leftrightarrow (\mat{B}-\mat{A})\in \posdef$. These matrices have real eigenvalues, which we can think of as axis lengths of ellipsoids. Order is given by ellipsoid inclusion.
\end{remark}

\begin{lemma}
The optimal control law for the LQG problem in \cref{def:ctlqg} is~$\bm{u}_t^\star =-\mat{K}\hat{\bm{x}}_t=-\mat{R}^{-1}\mat{B}^*\bar{\mat{S}}\hat{\bm{x}}_t$, where $\hat{\bm{x}}_t$ is the unbiased minimum-variance estimate of $\bm{x}_t$ given previous measurements and $\bar{\mat{S}}\in \posdef$ solves the Riccati equation
\begin{equation}
\label{eq:cont_ric_1}
    \mat{SA}+\mat{A}^*\mat{S}-\mat{SBR}^{-1}\mat{B}^*\mat{S}+\mat{Q}=\mat{0}.
\end{equation}
The minimum cost $J^\star$ achieved by the optimal control is\footnote{Note that~\cite{davis1977} contains a typo at p.188 (one extra $\mat{V}^{-1}$ factor). Instead,~\cite{kwakernaak1972} has a cleaner derivation and exposition.}:
\begin{equation*}
\label{eq:contmincost}
\begin{aligned}
J^\star&=\trace{\bar{\mat{S}}\bar{\mat{\Sigma}}\mat{C}^*\mat{V}^{-1}\mat{C}\bar{\mat{\Sigma}}+\bar{\mat{\Sigma}}\mat{Q}}\\
&=\trace{\bar{\mat{\Sigma}}\bar{\mat{S}}\mat{BR}^{-1}\mat{B}^*\bar{\mat{S}}+\bar{\mat{S}}\mat{W}},
\end{aligned}
\end{equation*}
where $\bar{\mat{\Sigma}}\in \posdef$ is the solution of the Riccati equation
\begin{equation}
\label{eq:cont_ric_2}
    \mat{A\Sigma} + \mat{\Sigma A}^*-\mat{\Sigma C}^*\mat{V}^{-1}\mat{C\Sigma}+\mat{W}=\mat{0}.
\end{equation}
\end{lemma}

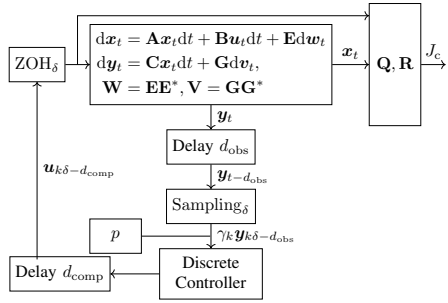
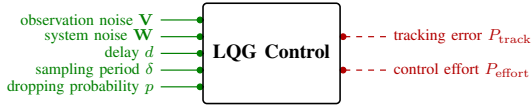
\begin{figure}[tb]
\begin{subfigure}{\columnwidth}
\begin{center}
\scalebox{0.65}{
    \begin{tikzpicture}
    \node[block] (dyn) at (0,0) {$
    \begin{aligned}
    \D \bm{x}_t&=\mat{A}\bm{x}_t\D t+\mat{B}\bm{u}_t \D t+\mat{E} \D \bm{w}_t\\
    \D \bm{y}_t &= \mat{C}\bm{x}_t \D t+\mat{G}\D \bm{v}_t,\\
    \mat{W}&=\mat{E}\mat{E}^*, \mat{V}=\mat{G}\mat{G}^*
    \end{aligned}$};
    \node[block, left=0.5cm of dyn.west] (zoh) {$\mathrm{ZOH}_\delta$};
    \node[block, below=0.5cm of dyn.south] (del_1) {Delay $d_\mathrm{obs}$};
    \node[block, below=0.5cm of del_1.south] (sam) {$\mathrm{Sampling}_\delta$};
    \node[block, below left=0.25cm and 0.5cm of sam.west] (gen) {$p$};
    \node[block, below=0.5cm of sam.south] (cnt) {\begin{tabular}{c}Discrete \\ Controller\end{tabular}};
    \node[block, right=0.75cm of dyn.east,minimum height=2.5cm] (cost) {$\mat{Q},\mat{R}$};
    \node[block, left=1cm of cnt.west] (del_2) {Delay $d_\mathrm{comp}$};
    \draw[->] ($(dyn.south)+(0,0)$) -- ($(del_1.north)$) node[pos=0.5,right]{$\bm{y}_t$};
    \draw[->] ($(del_1.south)+(0,0)$) -- ($(sam.north)$) node[pos=0.5,right]{$\bm{y}_{t-d_\mathrm{obs}}$};
    \draw[->] ($(dyn.east)+(0,0)$) -- ($(cost.west)$) node[pos=0.5,above]{$\bm{x}_t$};
    \draw[->] ($(zoh.east)+(0,0)$) -- ($(dyn.west)$) node[pos=0.5,below]{};
    \draw[->] ($(sam.south)+(0,0)$) -- ($(cnt.north)+(0,0)$) node[pos=0.5,right]{$\gamma_k \bm{y}_{k\delta-d_\mathrm{obs}}$};
    \draw[->] ($(cnt.west)+(0,0)$) -- ($(del_2.east)+(0,0)$) node[pos=0.5,right]{};
    \draw[-] ($(gen.east)+(0,0)$) -- ($(gen.east)+(1.4,0)$) node[pos=0.5,above]{};
    \draw[->] ($(zoh.south)+(0,-3.63)$) -- ($(zoh.south)+(0,0)$) node[pos=0.5,right]{$\bm{u}_{k\delta-d_\mathrm{comp}}$};
    \draw[->] ($(zoh.east)+(0.25,0)$) -- ($(zoh.east)+(0.25,1)$)|-($(cost.west)+(0,1)$) node[pos=0.96,above]{};
    \draw[->] ($(cost.east)$) -- ($(cost.east)+(0.5,0)$)node[pos=0.5,above]{$J_\mathrm{c}$};
    \end{tikzpicture}
    
\begin{comment}
    \begin{tikzpicture}
    \node[block] (dyn) at (0,0) {$\begin{aligned}
    \D \bm{x}_t&=\mat{A}\bm{x}_t \D t+\mat{B}\bm{u}_t\D t+\D \bm{w}_t\\
    \bm{y}_t&=\mat{C}\bm{x}_t \D t+ \D \bm{v}_t
    \end{aligned}$};
    \node[block, left=0.5cm of dyn.west] (zoh) {$\mathrm{ZOH}_\delta$};
    \node[block, below=0.75cm of dyn.south] (sam) {$\mathrm{Sampling}_\delta$};
    \node[block, below left=0.25cm and 1cm of sam.south] (cnt) {\begin{tabular}{c}Discrete \\ Controller\end{tabular}};
    \node[block, right=0.75cm of dyn.east,minimum height=2.5cm] (cost) {$\mat{Q},\mat{R}$};
    \draw[->] ($(dyn.south)+(0,0)$) -- ($(sam.north)$) node[pos=0.5,right]{$\bm{y}_t$};
    \draw[->] ($(dyn.east)+(0,0)$) -- ($(cost.west)$) node[pos=0.5,above]{$\bm{x}_t$};
    \draw[->] ($(zoh.east)+(0,0)$) -- ($(dyn.west)$) node[pos=0.5,below]{$\bm{u}_t$};
    \draw[->] ($(sam.south)+(0,0)$) -- ($(cnt.east)+(1,0)$)|-($(cnt.east)$) node[pos=0.5,right]{$\bm{y}_{k\delta}$};
    \draw[->] ($(cnt.west)$) -- ($(zoh.south)+(0,-2.5)$)-|($(zoh.south)$) node[pos=0.8,right]{$\bm{u}_{k\delta}$};
    \draw[->] ($(zoh.east)+(0.25,0)$) -- ($(zoh.east)+(0.25,0.85)$)|-($(cost.west)+(0,0.85)$) node[pos=0.96,above]{$\bm{u}_{t}$};
    \draw[->] ($(cost.east)$) -- ($(cost.east)+(0.5,0)$)node[pos=0.5,above]{$J_\mathrm{c}$};
    \end{tikzpicture}
\end{comment}}
\subcaption{LQG digital control with observation and computation delays, sampling and ZOH.}
\label{fig:digitallqg}
\end{center}
\end{subfigure}
\\[+5pt]
\begin{subfigure}{\columnwidth}
\begin{center}
\scalebox{0.75}{\begin{tikzpicture}[DP, dp port sep = 1.5]
            \node[dp={5}{2}] (cnt) {LQG Control};
            \draw[runconn, runame={tracking error $\track$}, relres=right] (cnt_res1){};
            \draw[runconn, runame={control effort $\effort$},relres=right] (cnt_res2){};
            \draw[funconn, funame={observation noise $\mat{V}$},relfun=left] (cnt_fun1){};
            \draw[funconn, funame={system noise $\mat{W}$},relfun=left] (cnt_fun2){};
            \draw[funconn, funame={delay $d$},relfun=left] (cnt_fun3){};
            \draw[funconn, funame={sampling period $\delta$},relfun=left] (cnt_fun4){};
            \draw[funconn, funame={dropping probability $p$},relfun=left] (cnt_fun5){};
\end{tikzpicture}}
\subcaption{Diagrammatic representation of co-design theorems.}
\label{fig:codesigntheorems}
\end{center}
\end{subfigure}
\caption{(a) We consider the case of LQG control first in the continuous-time and then in the digital case. We highlight the monotonicity in several nuisances in the problem. For instance, if the probability of dropping observations increases, the tracking error will not decrease. These properties are summarized diagrammatically in (b), which is the graphical representation of the feasibility relations presented in this work.}
\vspace{-0.5cm}
\end{figure}

\paragraph*{Co-design formalization}
To formalize the LQG control problem as a \gls{abk:dpi}, we first define two performance metrics. From a continuous-time LQG problem, we define the stationary \emph{tracking error} $\track$ and \emph{control effort} $\effort$:
\begin{equation*}
\track=\textstyle{\lim_{t\to \infty}}\mathbb{E}\{\bm{x}_t^\intercal \mat{Q}\bm{x}_t\}, \ \effort=\textstyle{\lim_{t\to \infty}}\mathbb{E}\{\bm{u}_t^\intercal \mat{R} \bm{u}_t \}.
\end{equation*}
We are now ready to state the first central result of this work.
\begin{theorem}
\label{thm:firstct}
The LQG problem of \cref{def:ctlqg} can be formulated and solved as a \gls{abk:dpi} with diagrammatic form as in~\cref{fig:dp_ct}.
\end{theorem}
To prove \cref{thm:firstct}, we show that there exists a \gls{abk:dpi} as in \cref{def:dpi}, relating \Ftext{functionalities} $\F{\mat{V}}, \F{\mat{W}}$ and \Rtext{resources} $\R{\track}, \R{\effort}$. First, by writing $\R{\track}$ and $\R{\effort}$ explicitly (\cref{lemma:precision}), we prove their monotonicity with respect to cost weighting (\cref{lemma:codesigncnt_1}). Second, we show the monotone relation characterizing the \gls{abk:dpi} (\cref{lemma:cont_V_W} and \cref{lemma:codesigncnt_2}).

\begin{lemma}
\label{lemma:precision}
The metrics $\track$ and $\effort$ can be written as
\begin{equation*}
\begin{aligned}
    \textstyle{\lim_{t\to \infty}}\mathbb{E}\{ \bm{x}_t^\intercal \mat{Q}_0\bm{x}_t\}&=\trace{\mat{Q}_0\left( \mat{\Sigma} +\mat{F}\right)}, \\
    \textstyle{\lim_{t\to \infty}}\mathbb{E}\{ \bm{u}_t^\intercal \mat{R}_0\bm{u}_t\}&=\trace{\mat{S}\mat{B}^*\mat{R}^{-1}\mat{R}_0\mat{R}^{-1}\mat{B}\mat{S}\mat{F}},
\end{aligned}
\end{equation*}
where 
$\mat{\Sigma}$ solves \cref{eq:cont_ric_2}, $\mat{F}$ solves the Lyapunov equation
\begin{equation}
\label{eq:lyapct}
    \left( \mat{A}-\mat{B}\mat{K}\right)\mat{F}+\mat{F}\left(\mat{A}-\mat{B}\mat{K}\right)^*+\mat{L}\mat{V}\mat{L}^*=\mat{0},
\end{equation}
$\mat{S}$ solves \cref{eq:cont_ric_1} and $\mat{L}=\mat{\Sigma}\mat{C}^*\mat{V}^{-1}$ is the Kalman gain.
\end{lemma}

Given explicit forms, we can now show that they are characterized by monotonic relations.
\begin{lemma}
\label{lemma:codesigncnt_1}
Let~$\mat{Q}(\alpha)=\alpha \mat{Q}_0$ and~$\mat{R}(\alpha)=\frac{1}{\alpha}\mat{R}_0$, $\alpha\in \mathbb{R}_+$. Let~$\bm{u}^\star(\alpha)$ be the solution of the LQG problem with $\mat{Q}(\alpha)$ and $\mat{R}(\alpha)$. Then, under optimal control one has:
\begin{compactitem}
    \item $P_\mathrm{track}(\alpha)$ is decreasing with $\alpha$
    increasing.
    \item $P_\mathrm{effort}(\alpha)$ is increasing with $\alpha$ increasing.
\end{compactitem}
\end{lemma}
\begin{comment}
\begin{proof}
First of all, notice that for~$\alpha\leq \alpha'$,~$\mat{Q}=\alpha\mat{Q}_0\preceq \alpha'\mat{Q}_0=\mat{Q}'$ and~$\mat{R}=\mat{R}_0/\alpha\succeq \mat{R}'=\mat{R}_0/\alpha'$. For the first part, following \cref{lemma:precision}, we need to prove that
\begin{equation*}
\label{eq:first_part}
    \alpha\leq \alpha' \Rightarrow \trace{\mat{Q}_0\left( \mat{\Sigma} +\mat{F}\right)}\mid_{\alpha}\geq \trace{\mat{Q}_0\left( \mat{\Sigma} +\mat{F}\right)}\mid_{\alpha'}.
\end{equation*}
$\mat{\Sigma}$ is independent from~$\alpha$. From \cref{eq:lyapct} it is clear that~$\alpha\leq \alpha' \Rightarrow \mat{F}_1\succeq \mat{F}_2$, which confirms the statement. For the second part, following \cref{lemma:precision}, it is easy to prove that $\alpha\leq \alpha'$ implies
%
%
$\trace{ \mat{S}\mat{B}\alpha^2\mat{R}_0^{-1}\mat{B}^*\mat{S}\mat{F}}\mid_{\alpha} \leq \trace{ \mat{S}\mat{B}\alpha'^2\mat{R}_0^{-1}\mat{B}^*\mat{S}\mat{F}}\mid_{\alpha'}.$
%
\vspace{-3mm}
\end{proof}
\end{comment}

Intuitively, by increasing $\alpha$ we increase the penalty for the tracking error of the control. For this reason, $\track$ decreases and $\effort$ increases. We now want to assess the effect of the system and observation noises on the optimal control.

\begin{lemma}
\label{lemma:cont_V_W}
The solution of  \cref{eq:cont_ric_2} is monotonic in $\mat{V}$ and $\mat{W}$, i.e., $\tup{\mat{V},\mat{W}}\preceq \tup{\mat{V}',\mat{W}'} \Rightarrow \mat{\Sigma}(\mat{V},\mat{W})\preceq \mat{\Sigma}(\mat{V}',\mat{W}')$.
\end{lemma}

\begin{comment}
\begin{proof}
First, we know that $\bar{\mat{\Sigma}}$ satisfies \cref{eq:cont_ric_2}:
\vspace{-0.1cm}
\begin{equation*}
    \mat{A\Sigma} + \mat{\Sigma A}^*-\mat{\Sigma C}^*\mat{V}^{-1}\mat{C\Sigma}+\mat{W}\coloneqq \mat{M}(\mat{V},\mat{W}).
    \end{equation*}
By letting $\tup{\mat{V}_1,\mat{W}_1}\preceq \tup{\mat{V}_2,\mat{W}_2}$, and using the notation $\mat{M}_i=\mat{M}(\mat{V}_i,\mat{W}_i)$, $i\in \{1,2\}$, from Theorem 3 in~\cite{de2004note} we know that $\mat{M}_2-\mat{M}_1 \succeq \mat{0} \Rightarrow \bar{\mat{\Sigma}}(\mat{V}_1,\mat{W}_1)\preceq \bar{\mat{\Sigma}}(\mat{V}_2,\mat{W}_2)$. We have:
\vspace{-0.2cm}
\begin{equation*}
        \mat{M}_2-\mat{M}_1=-\mat{\Sigma}\mat{C}^*\left( \mat{V}_2^{-1}-\mat{V}_1^{-1}\right)\mat{C}\mat{\Sigma}+\mat{W}_2-\mat{W}_1\succeq \mat{0},
\end{equation*}
where we use that $\mat{V}_1^{-1}-\mat{V}_2^{-1}\in \posdef$ and $\mat{W}_2-\mat{W}_1\in \posdef$ (\cref{lemma:invposef}).
Therefore, $\bar{\mat{\Sigma}}$ is monotone in $\mat{V}$ and $\mat{W}$.
\end{proof}
\end{comment}

\begin{lemma}
\label{lemma:codesigncnt_2}
Consider the situation of \cref{lemma:codesigncnt_1}:
\begin{compactitem}
    \item Fix $P_\mathrm{track}$. $P_\mathrm{effort}$ is monotonic in $\mat{W}$ and in $\mat{V}$.
    \item Fix $P_\mathrm{effort}$. $P_\mathrm{track}$ is monotonic in $\mat{W}$ and in $\mat{V}$.
\end{compactitem}
\end{lemma}
\begin{comment}
\begin{proof}
We need to prove:
\begin{equation*}
    \begin{aligned}
    \tup{\mat{V},\mat{W}}&\preceq \tup{\mat{V}',\mat{W}'}\Rightarrow \effort(\mat{V},\mat{W})\leq \effort(\mat{V}',\mat{W}'),\\
    \tup{\mat{V},\mat{W}}&\preceq \tup{\mat{V}',\mat{W}'}\Rightarrow \track(\mat{V},\mat{W})\leq \track(\mat{V}',\mat{W}').
    \end{aligned}
\end{equation*}
From \cref{lemma:cont_V_W}, we know $\mat{\Sigma}(\mat{V},\mat{W})\preceq \mat{\Sigma}(\mat{V}',\mat{W}')$. Furthermore, since $\mat{F}$ is the solution of the Lyapunov equation \cref{eq:lyapct}, we know that $\mat{F}(\mat{V},\mat{W})\preceq \mat{F}(\mat{V}',\mat{W}')$. As no other term of $\effort$ or $\track$ depends on $\mat{V},\mat{W}$, we are done.
\end{proof}
\end{comment}
This shows that the more uncertain the observations and the system dynamics are, the larger the control effort and tracking error will be, and concludes the proof of~\cref{thm:firstct}.

The presented \gls{abk:dpi} precisely assesses the feasibility relation between \Rtext{control effort}, \Rtext{tracking error}, \Ftext{system noise}, and the \Ftext{observation noise}. This \gls{abk:dpi} can be manipulated by taking the ``op'' of a quantity (\cref{def:dpi}), and moving it on the other side of the diagram. For instance, we can think of the observation noise as resource, by switching its meaning to \Rtext{information} (i.e., from noise to information matrix).

\noindent \paragraph*{Dealing with delays} We now show the ability of our formalism to capture the influence of delays on the system.

\begin{theorem}
\label{thm:delay}
A continuous-time LQG problem with observation and computation delays ($d_\mathrm{obs}$, $d_\mathrm{comp}$) can be formulated and solved as a \gls{abk:dpi} with diagrammatic as in \cref{fig:dp_ct_del}.
\end{theorem}
To establish the effect of a nuisance, we follow what we call the \emph{substitution principle}. If in the case in which the nuisance was ``lower'' the controller could simulate a ``higher'' nuisance, then we have monotonicity. If we had a smaller delay, we could simulate a larger one by adding it artificially. Hence, \Rtext{control effort} and \Rtext{tracking error} of the optimal control strategy cannot decrease with larger \Ftext{delay}.

\paragraph*{Visualization via Pareto fronts}
We want to give a visual interpretation of the presented results. For the scalar case of \cref{def:ctlqg} we can derive $\track$ and $\effort$ in  closed-form:
\begin{equation*}
    \begin{aligned}
    \track(q_0) &=q_0\left( \bar{\sigma}+\frac{\bar{(\sigma}c)^2}{2v\sqrt{a^2+\alpha^2q_0b^2/r_0}}\right),\\
    \effort(r_0)&=\frac{r_0\bar{\sigma}^2c^2}{2b^2v}\frac{\left(a+\sqrt{a^2+\alpha^2b^2q_0/r_0}\right)^2}{\sqrt{a^2+\alpha^2b^2q_0/r_0}},
    \end{aligned}
\end{equation*}
where variables are in lower case since they represent scalar quantities. We can compute their limits, by fixing $v$ and $w$:
\begin{equation*}
\begin{aligned}
\lim_{\alpha \to 0}\track(q_0)&=q_0\left(\bar{\sigma}+\frac{(\bar{\sigma}c)^2}{2va}\right),\lim_{\alpha \to \infty}\track(q_0)=q_0\bar{\sigma},\\
\lim_{\alpha \to 0}\effort(r_0)&=\frac{2r_0a(c\bar{\sigma})^2}{{b^2v}},
\lim_{\alpha \to \infty}\effort(r_0)= \infty .
\end{aligned}
\end{equation*}
We can plot instances of the Pareto front $\tup{\R{\track},\R{\effort}}$ (\cref{fig:pareto}). This is the query in which we ``fix functionalities and minimize resources'', where the Pareto fronts represent the best achievable control performances, given specific $v$ and $w$. Alternatively, we could ask to ``fix resources and maximize functionalities''. This is the query in which we fix $\R{\track}$ and $\R{\effort}$, and observe a Pareto front of system and observation noises at which the given performance can be provided. Monotonicity is easy to see. By choosing $\tup{\F{v_1},\F{w_1}}\preceq \tup{\F{v_2},\F{w_2}}$ (i.e., $\F{v_1}\leq \F{v_2}$ and $\F{w_1}\leq \F{w_2}$), we see that both~$\R{\track}$ and~$\R{\effort}$ increase, and hence that the Pareto fronts dominate each other.
\section{Digital LQG control as a \gls{abk:dpi}}
\label{sec:part_ii} 
\paragraph*{Digital LQG control} We define the infinite-horizon discrete-time LQG control problem.
\begin{definition}[Discrete-time LQG control]
\label{def:dtlqg}
Consider the discrete-time stochastic dynamics
\begin{equation*}
\label{eq:cont_dyn}
    \begin{aligned}
    \bm{x}_k&=\dis{\mat{A}}\bm{x}_k+\dis{\mat{B}}\bm{u}_k +\dis{\mat{E}} \bm{w}_k\\
    \bm{y}_k &= \dis{\mat{C}}\bm{x}_k + \dis{\mat{G}}\bm{v}_k,
    \end{aligned}
\end{equation*}
where $\bm{w}_k$ and $\bm{v}_k$ are two standard Brownian processes and $\dis{\mat{W}}=\mat{E}\mat{E}^*$, $\dis{\mat{V}}=\mat{G}\mat{G}^*$ the noise covariances. The \emph{discrete-time} infinite-horizon LQG problem consists of finding a control law $\bm{u}^\star$ which minimizes the quadratic cost
\begin{equation*}
\label{eq:general_discrete_cost}
    \dis{J}=\textstyle{\lim_{N\to \infty}}\frac{1}{N}\mathbb{E}\{ \sum_{i=0}^{N-1}\left( \bm{x}_i^\intercal \dis{\mat{Q}}\bm{x}_i+\bm{u}_i^\intercal \dis{\mat{R}}\bm{u}_i\right)\},
\end{equation*}
where $\dis{\mat{Q}}\in \possemidef$, $\dis{\mat{R}}\in \posdef$.
\end{definition}

We want to show that the relations found in \cref{sec:part_i} hold for the case of digital LQG control, where we want to control a continuous-time system using a digital controller.

\begin{lemma}
\label{lemma:discretization}
Consider a continuous-time LQG problem, where observations are sampled with period $\delta$ and processed by a digital controller to produce a control input (\cref{fig:digitallqg}). The input is reconstructed using ZOH with period $\delta$. We can find:
\begin{equation*}
\begin{aligned}
\dis{\mat{A}}&=e^{\mat{A}\delta}, \dis{\mat{B}}=\smallint_{0}^{\delta}e^{\mat{A}s}\mat{B}\D s, \dis{\mat{C}}=\mat{C},
\dis{\mat{Q}}=\smallint_{0}^\delta e^{\mat{A}^\intercal s}\mat{Q}e^{\mat{A} s}\D s,\\
\dis{\mat{R}}&=\smallint_{0}^{\delta} ( (\smallint_{0}^{s} e^{\mat{A}t}\mat{B} \D t) \mat{Q} ( \smallint_{0}^{s} e^{\mat{A}t}\mat{B} \D t) ^\intercal + \mat{R}) \D s.
\end{aligned}
\end{equation*}
such that the optimal cost of this controller coincides with the optimal cost $\dis{J}$ in \cref{def:dtlqg}.
\end{lemma}

\noindent Define the stationary digital \emph{tracking error} and \emph{control effort}:
\begin{equation*}
\track^{\mathrm{d}}=\lim_{k\to \infty}\mathbb{E}\{\bm{x}_k^\intercal \mat{Q}\bm{x}_k\},~
\effort^{\mathrm{d}}=\lim_{k\to \infty}\mathbb{E}\{\bm{u}_k^\intercal \mat{R} \bm{u}_k \}.
\end{equation*}
We can now formulate \cref{thm:firstct} for the discrete-time case.
\begin{theorem}
\label{thm:codesigndt}
The digital LQG problem in \cref{lemma:discretization} with fixed sampling period $\delta$ can be formulated and solved as a \gls{abk:dpi} with diagrammatic form as in~\cref{fig:dp_ct}.
\end{theorem}
The proof is analogous to the one of \cref{thm:firstct}.
\cref{thm:codesigndt} allows us to interconnect LQG control problems as \glspl{abk:dpi} in the co-design diagram of a real robotic platform.
\paragraph*{Effect of sampling period and intermittent observations} We now assess the impact of sampling period and intermittent observations~\cite{sinopoli2004,censi2011} on digital LQG control. 
\begin{definition}[LQG with intermittent observations]
\label{def:lqgintermittent}
The \emph{LQG problem with intermittent observations} differs from the original problem by the observations $\bm{y}_k'=\gamma_k\bm{y}_k$, where $\gamma_k\in \{0,1\}$ is a random sequence. 
\end{definition}

\begin{comment}
\begin{lemma}
Consider the LQG problem with intermittent observations given in \cref{def:lqgintermittent}. Denote by $p$ the probability that observation $\bm{y}_t$ is dropped. One has
\begin{itemize}
    \item Fix $\effort^\mathrm{d}$, $\track^\mathrm{d}$ is monotonic with $p$.
    \item Fix $\track^\mathrm{d}$, $\effort^\mathrm{d}$ is monotonic with $p$.
\end{itemize}
\end{lemma}
\end{comment}

\begin{theorem}
\label{thm:drop}
The digital LQG problem given in \cref{fig:digitallqg}
with delay of the form~$\delta=2^n\delta_0$ can be formulated and solved as a \gls{abk:dpi} with diagrammatic form as in \cref{fig:codesigntheorems}.
\end{theorem}
To prove both cases, we can use the substitution principle. Assuming a sampling period $\delta=2^n\delta_0$, $\track$ and $\effort$ are monotonic with $n$.\footnote{Note that using $\delta$ of this form corresponds to the case where the information available is a subset decreasing with~$n$.
We are not stating this result in general. This would require a deeper discussion and several assumptions about the system (e.g., about oscillatory behavior). This is an open problem~\cite{melzer1971,bini2013} and we leave its extensive treatment to future work.}
On the other hand, if the controller is given a set of observations, it can simulate having less (i.e., an higher drop probability $p$), by artificially deleting selected observations. Therefore, the control effort and tracking error cannot decrease with increasing $p$.

\paragraph*{Discussion}
\cref{thm:firstct,thm:delay,thm:codesigndt,thm:drop} show how to frame and solve variations of LQG control problems as \glspl{abk:dpi}. In particular, the results provide an interface to include control synthesis in the robot co-design problem. Note that the theory and results generalize to the case of parameter uncertainty~\cite{censi2017uncertainty}. 
In the next section we show how this new perspective allows one to solve complex robot co-design problems, capturing heterogeneous abstraction levels within a unified framework.
\section{Co-design of an autonomous drone}
\label{sec:case_study}
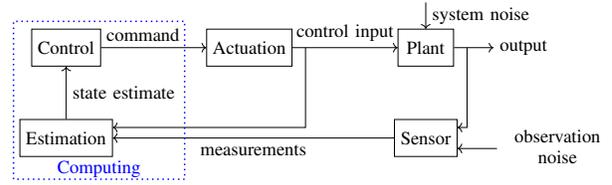
\begin{figure}[t]
\begin{center}
\scalebox{0.7}{\begin{tikzpicture}
\node[block] (cnt) at (0,0) {Control};
\node[block, right=2cm of cnt] (act) {Actuation};
\node[block, right=2cm of act] (plant) {Plant};
\node[block, below=1cm of cnt] (est) {Estimation};
\node[block, below=1cm of plant] (sen) {Sensor};
\draw[->] (cnt) -- (act) node[pos=0.4,above]{command};
\draw[->] (act) -- (plant) node[pos=0.5,above]{control input};
\draw[->] (sen) -- (est) node[pos=0.5,below]{measurements};
\draw[->] (est) -- (cnt) node[pos=0.5,right]{state estimate};
\draw[->] ($(sen.east)+(0.75,-0.2)$) -- ($(sen.east)+(0,-0.2)$) node[pos=0,right]{\begin{tabular}{c}observation\\ noise\end{tabular}};
\draw[->] ($(plant.east)+(0,0)$) -- ($(plant.east)+(0.75,0)$) node[pos=1,right]{output};
\draw[->] ($(plant.east)+(0.25,0)$) -- ($(plant.east)+(0.25,-1.5)$)|- ($(sen.east)+(0,0.2)$);
\draw[->] ($(act.east)+(0.25,0)$) -- ($(act.east)+(0.25,-1.5)$)|- ($(est.east)+(0,0.2)$);
\draw[->] ($(plant.north)+(0,0.5)$) -- ($(plant.north)+(0,0)$)node[pos=0.5,right]{system noise};
\draw[dotted,blue,thick] (-1,-2.5) rectangle (2.25,0.5) {};
\node[blue] at (0.625,-2.3) {Computing};

\end{tikzpicture}}
\caption{Data-flow diagram for an autonomous system. \label{fig:signalflow}}
\end{center}
\vspace{-0.6cm}
\end{figure}
In this section we show the ability of the proposed formalism to capture LQG control synthesis together with practical aspects related to robot design, such as sensor, actuation and computer selection. We consider a sensor-based navigation task for an autonomous drone (\cref{fig:drone_pics}). The drone must reach a goal and it is guided by a vision sensor, used as a bearing sensor, needed to detect the relative direction of the goal.
\subsection{System modeling}
To understand the co-design approach, we need to distinguish logical dependencies and data-flow. While the diagram provided in \cref{fig:signalflow} describes the system's data-flow, the co-design diagram in \cref{fig:drone_codesign} formalizes logical dependencies through functional decomposition. We now present the single \glspl{abk:dpi} composing the co-design diagram, together with an explanation on how to obtain the models.

\emph{Block \textbf{LQG Control}}: Given the task of aligning itself with the goal, we define the state of the robot as~$\tup{\theta_t,\omega_t}$, where $\theta_t$ is its heading and $\omega_t$ its angular speed. The heading of the goal at time $t$ is denoted by $\theta_t^\mathrm{g}$. The control is the resulting torque $\tau_t$. The dynamics of the heading are given by the differential equations~$\D \theta_t =\omega_t \D t$, $\D \omega_t = \frac{\tau_t}{I}\D t +\D w_t$, where $I$ is the moment of inertia of the drone about its rotational axis, and $w_t$ is a Brownian process with intensity $\F{\sigma_w^2}$. We assume Gaussian observations of the relative bearing of drone and goal~$y_t=(\theta_t-\theta_t^{\mathrm{g}})+v_t$, where $v_t$ is white Gaussian noise with intensity $\R{\sigma_v^2}$, describing measurement uncertainty. Assuming that the goal is far away, $\theta_t^\mathrm{g}$ is approximately constant, and w.l.o.g. we can assume it to be 0 ($y_t=\theta_t+v_t$). We pose the stationary problem of minimizing the objective $\textstyle{\lim_{T\to \infty}} \frac{1}{T} \smallint_{0}^T \left( \alpha q_0 \theta_t^2+\frac{r_0}{\alpha} \tau_t^2 \right) \D t.$ This is a LQG problem with $\mat{Q}=\text{diag}(0,\alpha q_0)$, $\mat{R}=r_0/\alpha$, for the continuous-time system given by the system matrices:
\begin{equation*}
    \begin{aligned}
        \mat{A}&=\begin{bmatrix}
        0&1\\
        0&0
        \end{bmatrix},\quad \mat{B}=\frac{1}{I}\begin{bmatrix} 0\\ 1\end{bmatrix},\quad  \mat{C}=\begin{bmatrix} 1&0\end{bmatrix},\\ \mat{W}&=\begin{bmatrix} 0&0\\ 0&\sigma_w^2/I\end{bmatrix},\quad \mat{V}=\sigma_v^2.
    \end{aligned}
\end{equation*}
As shown in~\cref{sec:part_i}, we can formalize this as a \gls{abk:dpi}~\circled{1}, which provides stability of the system up to a given \Ftext{system noise}, requiring \Rtext{observations} at a specific frequency and with given \Rtext{precision}. The control law has to be implemented at a given \Rtext{frequency}, resulting in specific \Rtext{control effort} and \Rtext{tracking error}. Implementations are given by the different cost weights, parametrized by $\alpha$. \emph{Obtaining the model}: As explained in \cref{sec:part_i}, the \gls{abk:dpi} can be obtained by solving specific Riccati equations via numerical simulations.

\emph{Block \textbf{Vision Sensor}}: The observations required by the control block are provided by a \textbf{vision sensor} \circled{6} at a given \Ftext{frequency} and with a specific \Ftext{resolution}. Such sensors have a \Rtext{cost}, \Rtext{mass} and \Rtext{power consumption}. \emph{Obtaining the model}: This is obtained from sensor catalogues (e.g., cameras).

\emph{Block \textbf{Feature Extraction}}: Sensor measurements are processed by a \textbf{feature extraction} algorithm~\circled{3}, providing the LQG control \gls{abk:dpi} with observations at a certain \Ftext{frequency} and \Ftext{accuracy}, limited by sensor properties.
\emph{Obtaining the model}: This can be obtained by answering photogrammetry questions such as ``what \Rtext{resolution} is needed to achieve a certain detection \Rtext{accuracy}?''

\emph{Blocks \textbf{Algorithms Implementation}}: Given the above, we need to implement the \textbf{control and feature detection algorithms}~\circled{8}, \circled{9}. Each of these is a \gls{abk:dpi}, characterized by a catalogue of algorithms, each operating at a specific \Ftext{frequency} and requiring \Rtext{computation power}. \emph{Obtaining the model}: This can be obtained via benchmarking. An example is given by \emph{SLAMBench}~\cite{nardi15}.

\emph{Block \textbf{Computing}}: The processes require \Ftext{computation power}, provided by a \textbf{computing unit}~\circled{5}. Each computing unit has a \Rtext{cost}, \Rtext{mass}, and requires \Rtext{power}. \emph{Obtaining the model}: We can model this via computer catalogues.

\emph{Block \textbf{Actuation}}: The total mass of the system is lifted thanks to \textbf{actuation}~\circled{2}, which provides \Ftext{lift}, \Ftext{control effort} and \Ftext{speed} by requiring \Rtext{cost}, \Rtext{mass} and \Rtext{power}.
\emph{Obtaining the model}: Models can be obtained from catalogues. For instance, power consumption can be modeled as a monotone function of \Ftext{lift} and \Ftext{control effort}, as in~\cite{censi2015}.

\emph{Block \textbf{Battery}}: A \textbf{battery}~\circled{7} provides \Ftext{energy} to the system, requiring \Rtext{cost} and \Rtext{mass}. \emph{Obtaining the model}: Different battery technologies can be extracted from catalogues~\cite{censi2015}.

\emph{Block \textbf{Mission Planning}}: Finally, a \textbf{mission planning} block \circled{4} evaluates the performance of the system, measured by \Ftext{tracking error}, the \Ftext{mission time}, the \Ftext{number of missions} and the detected \R{drone speed}. \emph{Obtaining the model}: This is a simple list of requirements for specific scenarios. 

\begin{comment}
\begin{figure}[tb]
\begin{subfigure}[b]{0.49\columnwidth}
\begin{center}
%
\input{ACC/tikz/computer.tikz}
\caption{Computer design problem.\label{fig:computer}}
\end{center}
\end{subfigure}
\begin{subfigure}[b]{0.49\columnwidth}
\begin{center}
%
\input{ACC/tikz/camera.tikz}
\caption{Vision sensor design problem.\label{fig:vision}}
\end{center}
\end{subfigure}
~
\begin{subfigure}[b]{0.49\columnwidth}
\begin{center}
%
\input{ACC/tikz/feature.tikz}
\caption{Feature extraction design problem.\label{fig:feature}}
\end{center}
\end{subfigure}
\begin{subfigure}[b]{0.49\columnwidth}
\begin{center}
%
\input{ACC/tikz/battery.tikz}
\caption{Battery design problem.\label{fig:battery}}
\end{center}
\end{subfigure}
~
\begin{subfigure}[b]{0.49\columnwidth}
\begin{center}
%
\input{ACC/tikz/thrust.tikz}
\caption{Actuation design problem.\label{fig:actuation}}
\end{center}
\end{subfigure}
\begin{subfigure}[b]{0.49\columnwidth}
\begin{center}
%
\input{ACC/tikz/efficiency.tikz}
\caption{Mission planning design problem. \label{fig:mission}}
\end{center}
\end{subfigure}
\caption{Components of the drone co-design problem.}
\label{fig:components}
\end{figure}
\end{comment}

\paragraph*{Composing the full diagram}
The single \glspl{abk:dpi} presented in the previous paragraph are interconnected to form a co-design diagram (\cref{fig:drone_codesign}) as follows. First, the total power required by the system arises from the sum of the power required by actuation, by the sensors and by the computing unit. Given the \Ftext{mission time}, one can determine the \Ftext{energy} which needs to be provided by the battery. This is the first feedback loop in the co-design diagram. Second, the computation required by both the control and feature extraction implementations needs to be provided by the computing unit (second loop). Third, the mass of the system is given by the masses of the sensors, battery, actuators and computing unit, and determines the \Ftext{lift} needed from actuation (third loop). The number of loops provides an upper bound for the complexity of the solution algorithm~\cite[Proposition 5]{censi2015}. In particular, the solver will create a chain in the poset of antichains of $\bar{\mathbb{R}}_{\geq 0}^3$, where the latter represents the \gls{abk:poset} containing the three resources which are fed back.

\subsection{Co-Design results}
\begin{figure}[t]
\begin{subfigure}[b]{\columnwidth}
\begin{center}
\includegraphics[width=\linewidth]{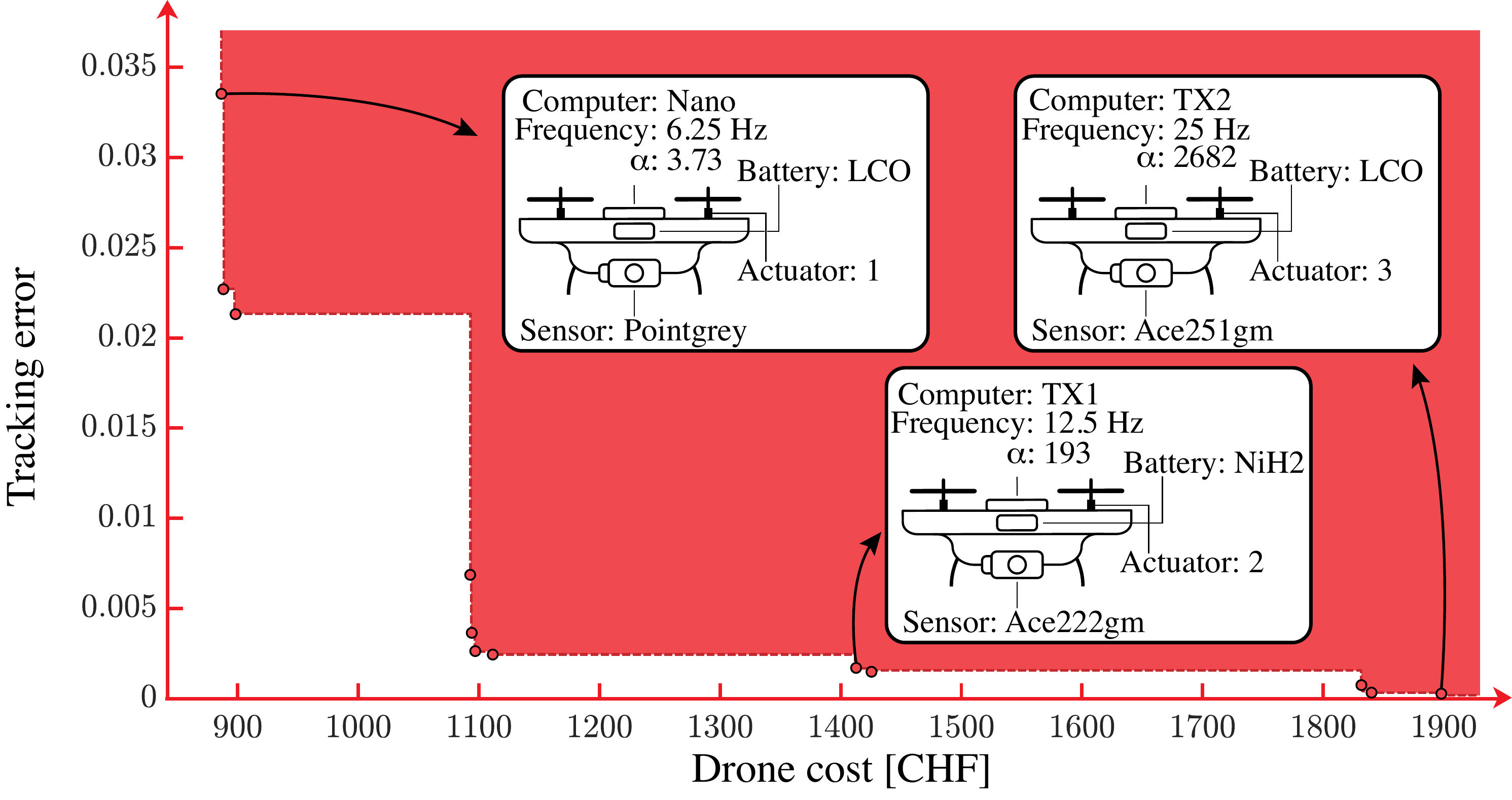}
\end{center}
\subcaption{Pareto front of \Rtext{cost} and \Rtext{tracking error} (performance) in the design of a drone, able to complete 5,000 missions of 40 minutes.}
\label{fig:res_1}
\end{subfigure}
\begin{subfigure}[b]{\columnwidth}
\begin{center}
\includegraphics[width=\linewidth]{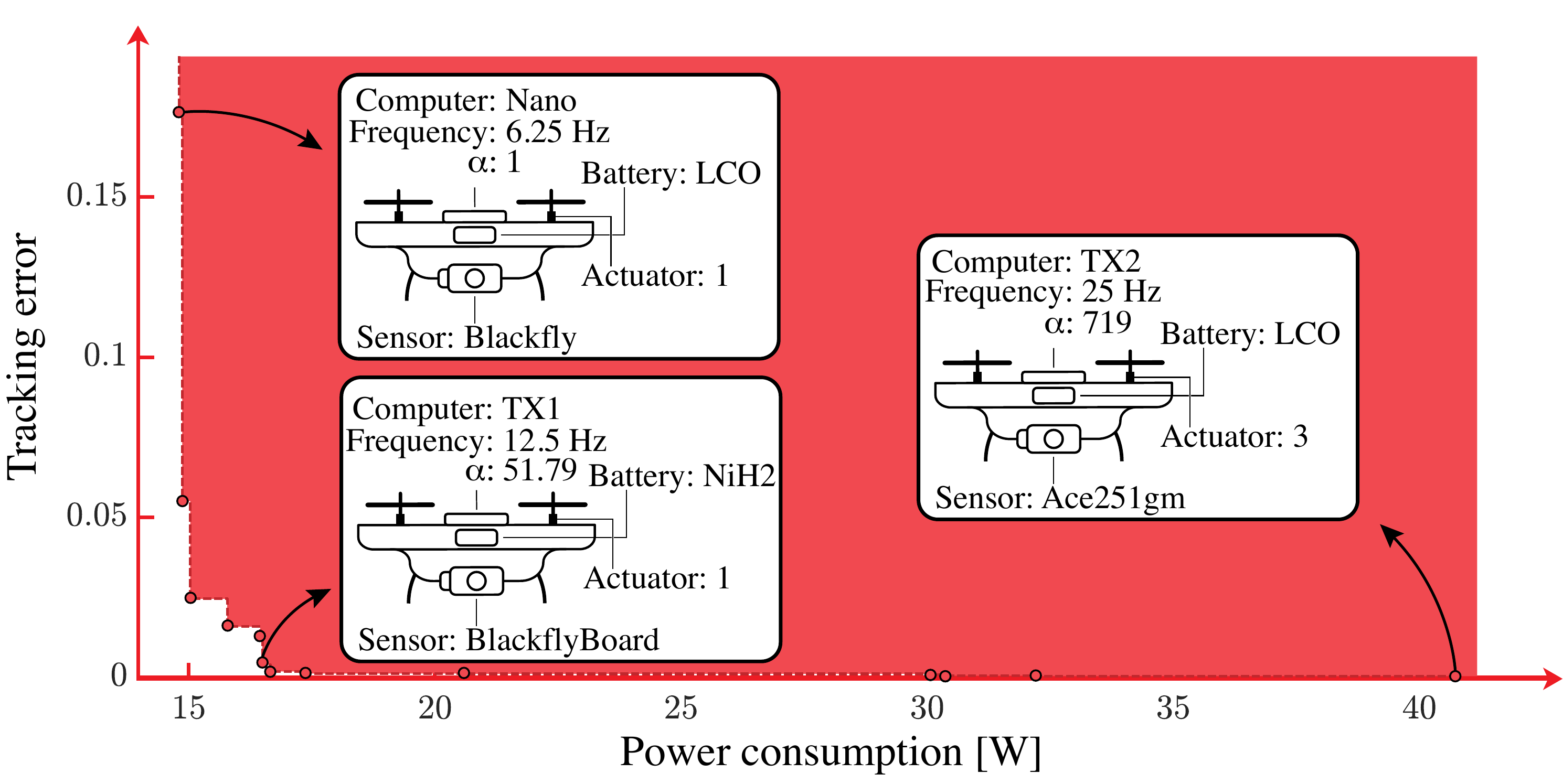}
\end{center}
\subcaption{Pareto front of \Rtext{power consumption} and \Rtext{tracking error} (performance) in the design of a drone, able to complete 1,000 missions of 10 minutes.}
\label{fig:res_2}
\end{subfigure}
\begin{subfigure}[b]{\columnwidth}
\begin{center}
\includegraphics[width=\linewidth]{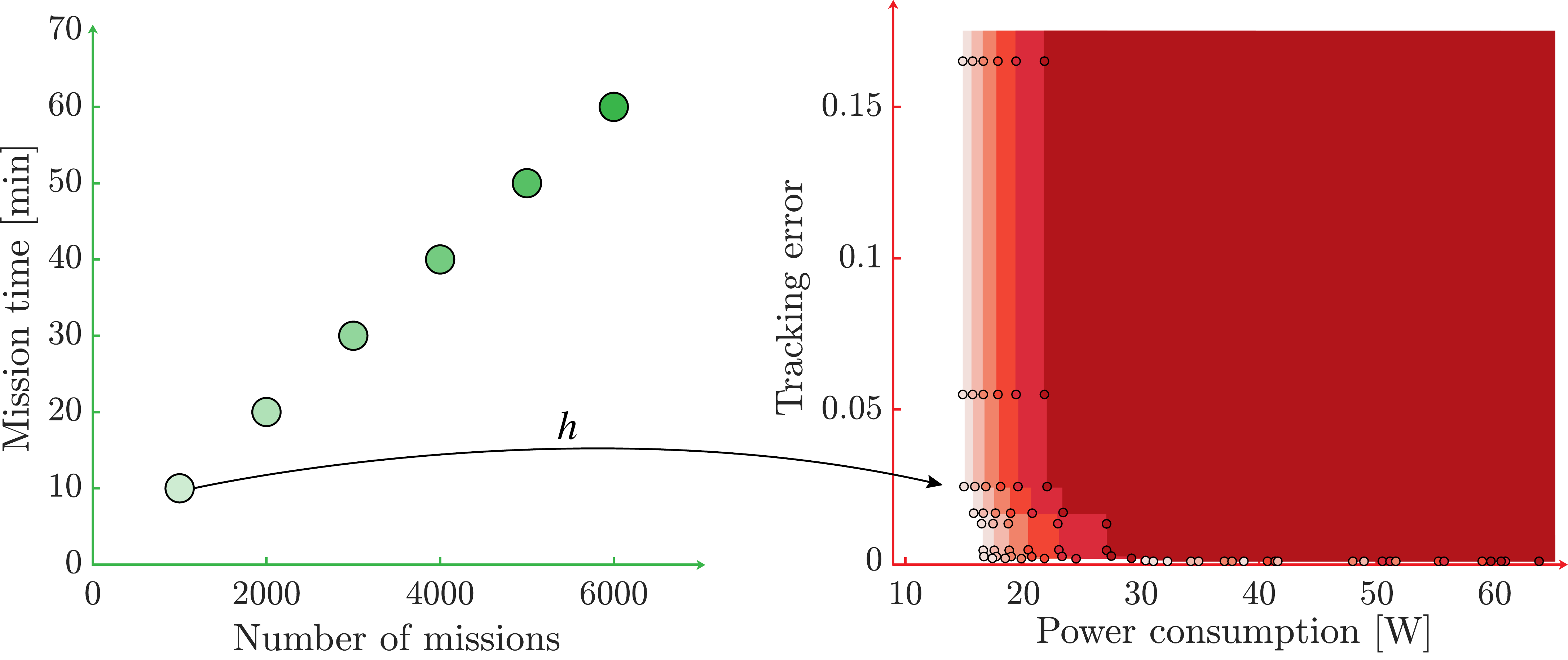}
\end{center}
\subcaption{Monotonicity of the drone \gls{abk:dpi}. Higher \Ftext{mission time} and \Ftext{number of missions} requires higher \Rtext{power} and \Rtext{tracking error}.}
\label{fig:res_3}
\end{subfigure}
\caption{Trade-offs of \Ftext{mission time}, \Ftext{number of missions}, \Rtext{cost}, \Rtext{tracking error} and \Rtext{power consumption} for the co-design of an autonomous drone. (a-b) The figure shows the antichain of optimal solutions for a given scenario. Red dots characterize optimal design solutions and the colored area describes upper-sets of resources for which functionalities are feasible. Selected implementations corresponding to specific points in the antichain are reported. (c) Pareto fronts of resources as a function of the (increasing) functionalities. The dominance of subsequent Pareto fronts describes monotonicity.}
\label{fig:results_first}
\end{figure}
We now show how the proposed framework is able to solve the co-design problem of an autonomous drone, co-designing the controller synthesis together with the rest of the platform. We consider the \gls{abk:dpi} reported in \cref{fig:drone_codesign} and provide design insights in terms of \Rtext{cost}, \Rtext{power consumption}, \Rtext{tracking error}, \Ftext{missions duration} and \Ftext{number of missions}.\footnote{The solver, based on a formal language, is available at \href{codesign.science}{codesign.science}.} We list the design variables in \cref{tab:designvars}. Specifically, they include the selection of sensors, control parameters, battery technologies, computing units and actuators. In the following, we bound the \Ftext{system noise} and fix $q_0=r_0=1$, to propose a sample of design insights that the framework can produce.
\begin{remark}
The following results have to be understood qualitatively, not quantitatively: we contribute in the formalization of meta-models and interfaces, not in the  numbers. We look forward to working with experts to get adequate data. 
\end{remark}
\paragraph*{Cost and performance trade-off (\cref{fig:res_1})}
We query the optimal design solutions which enable the drone to perform 5,000 missions lasting 40 minutes (\cref{fig:res_1}). The red dots represent the elements belonging to the antichain of optimal design solutions, expressed in terms of the \Rtext{platform cost} and the \Rtext{tracking error}. The solutions are not comparable, since no instance leads simultaneously to lower \Rtext{cost} and \Rtext{tracking error}. The upper set of resources (not necessarily optimal, but making the \gls{abk:dpi} feasible) is given in solid red. We report implementations corresponding to specific optimal solutions. As can be gathered from \cref{fig:res_1}, a budget increase for the drone reduces the tracking error. For instance, an investment from \unit[900]{CHF} to \unit[1,100]{CHF} reduces the tracking error by \unit[90]{\%}. On the other hand, a \unit[27]{\%} investment from \unit[1,100]{CHF} to \unit[1,400]{CHF} only reduces the tracking error by \unit[5]{\%}.
\paragraph*{Power and performance trade-off (\cref{fig:res_2})}
The optimization objectives can be quickly modified, without affecting the way the solution is obtained. For instance, we query the co-design framework in a similar way, now looking at the trade-offs between \Rtext{power consumption} and \Rtext{tracking error} for the case in which the drone must complete 1,000 missions lasting 10 minutes (\cref{fig:res_2}). When more power is available, better sensors and more performing computers, batteries and actuators can be used, reducing the tracking error.

\paragraph*{Monotonicity of the drone \gls{abk:dpi} (\cref{fig:res_3})}
We consider increasing \Ftext{mission time} and \Ftext{number of missions} and assess the evolution in trade-offs in platform \Rtext{power consumption} and \Rtext{tracking error}. \cref{fig:res_3} shows multiple co-design queries. In particular, for each \Ftext{functionality} (left plot), we compute the map $h$, which maps a \Ftext{functionality} to the minimum antichain of \Rtext{resources} which provide it (\cref{def:map_h}). Monotonicity can be seen in the dominance of subsequent Pareto fronts (right plot), illustrated in increasing red tonality.

\setlength{\tabcolsep}{3pt}
\begin{table}[t]
\begin{footnotesize}
\begin{center}
\begin{tabular}{lll}
\textbf{Variable} & \textbf{Options}& \textbf{Source}\\
\toprule
Actuators& Type 1, 2, 3 &\cite{censi2015}\\
Computer&RaspPi 4B, Jetson Nano/TX1/TX2&\cite{nvidia}\\
Control& \unit[0.2-50.0]{Hz}, $\alpha \in [10^{-4},10^4]$&-\\
Sensor &Basler Ace251gm/222gm/13gm/7gm/5gm/15um & \cite{basler}\\
& Flir Pointgrey/Blackfly/BlackflyBoard&\cite{flir}\\
Battery &LCO,~LFP,~LiPo,~LMO,~NiCad,~NiH2,~NiMH,~SLA&\cite{censi2015}\\
\bottomrule
\end{tabular}
\end{center}
\caption{Variables, options and sources for the drone co-design problem.}
\label{tab:designvars}
\end{footnotesize}
\end{table}
\section{Conclusions}
\label{sec:conclusion}
We considered the co-design of robotic platforms and linear control systems steering them. We have shown how to frame LQG control problems in a monotone theory of co-design, framing them as monotone feasibility relations (\glspl{abk:dpi}). The approach guarantees modularity and compositionality, allowing us to model complex, interconnected systems and to capture heterogeneous design abstraction levels, ranging from control synthesis to hardware selection.
\paragraph*{Outlook}
This work fits in our vision of compositional robotics, where we want to describe and co-design robots all the way from signals characterising a single component~\cite{zardini2020compositional} to networks and fleet effects~\cite{ZardiniEtAlBis2020}. We want to leverage compositionality properties of our framework to interconnect the co-design problem of a robot in the co-design problem of an urban robot-enabled mobility system, to assess how local trade-offs influence fleet-level operations~\cite{choudhury2020efficient}. Furthermore, we want to extend the proposed studies to the synthesis of more complex control structures.

\section*{Acknowledgments}
We thank Dr. Fuller, Dr. Bonalli, Dr. Tanaka, Mr. Lanzetti, Mr. Zanardi, and Ms. Monti.
\appendix
\label{sec:appendix}
\subsection{Background on orders}
\label{sec:app_order}
\begin{definition}[Poset]
A \emph{\gls{abk:poset}} is a tuple $\langle P,\preceq_P\rangle$, where $P$ is a set and $\preceq_P$ is a partial order, defined as a reflexive, transitive, and antisymmetric relation.
\end{definition}

\begin{comment}
\begin{definition}[Poset of Hermitian matrices]
We introduce the Hermitian matrices poset $\tup{\mathcal{M},\preceq}$, which is instrumental in this article.
Given two Hermitian matrices $\mat{A},\mat{B}\in \mathcal{M}$, one has $\mat{A}\preceq \mat{B} \Leftrightarrow (\mat{B}-\mat{A})\in \posdef$. Since the matrices are Hermitian, they are guaranteed to have real eigenvalues. Therefore, we can think of the eigenvalues as axis lengths of ellipsoids, and we can order matrices by ellipsoid inclusion.
\end{definition}
\end{comment}

\begin{definition}[Opposite of a poset]
The \emph{opposite} of a poset $\langle P,\preceq_P \rangle$ is the poset $\langle P^\mathrm{op},\preceq_P^\mathrm{op} \rangle$, which has the same elements as $P$, and the reverse ordering.
\end{definition}
\begin{definition}[Product poset]
Let $\tup{P,\preceq_{P}}$ and $\tup{Q,\preceq_{Q}}$ be posets. Then,~$\tup{P\times Q,\preceq_{P\times Q}}$ is a poset with
\begin{equation*}
    \tup{p_1,q_1}\preceq_{P\times Q}\tup{p_2,q_2} \Leftrightarrow (p_1\preceq_{P}p_2) \wedge (q_1\preceq_Q q_2).
\end{equation*}
This is called the \emph{product poset} of $\tup{P,\preceq_{P}}$ and $\tup{Q,\preceq_{Q}}$. 
\end{definition}

\begin{definition}[Monotonicity]
A map~$f\colon P\rightarrow Q$ between two posets~$\langle P, \preceq_P \rangle$,~$\langle Q, \preceq_Q \rangle$ is \emph{monotone} iff~$x\preceq_P y$ implies $f(x) \preceq_Q f(y)$. Monotonicity is compositional.
\end{definition}

\subsection{Proofs for \cref{sec:part_i}}
\label{sec:app_A}

\begin{proof}[Proof of \cref{lemma:precision}]
With $\bm{e}_t=\bm{x}_t-\hat{\bm{x}}_t$ estimation error one has: 
\begin{equation*}
    \begin{aligned}
        \lim_{t\to \infty}\mathbb{E}\{\bm{x}_t^\intercal \mat{Q}_0\bm{x}_t\}&=\lim_{t\to \infty}\trace{\mat{Q}_0\mathbb{E}\{\bm{x}_t\bm{x}_t^\intercal\}}\\
        &=\lim_{t\to \infty} \trace{\mat{Q}_0 \mathbb{E}\{(\hat{\bm{x}}_t+\bm{e}_t)(\hat{\bm{x}}_t+\bm{e}_t)^\intercal\}}\\
        &=\lim_{t\to \infty }\trace{\mat{Q}_0 \mathbb{E} \{\hat{\bm{x}}_t\hat{\bm{x}}_t^\intercal +\bm{e}_t\bm{e}_t^\intercal\}}\\
        &=\trace{\mat{Q}_0(\mat{F}+\mat{\Sigma})},
    \end{aligned}
\end{equation*}
where $\mat{F}$ can be computed using the closed-loop dynamics in \cref{eq:lyapct}. By applying the optimal control $\bm{u}_t^\star$, we have:
\begin{equation*}
    \begin{aligned}
        \lim_{t\to \infty}\mathbb{E}\{\bm{u}_t^\intercal \mat{R}_0\bm{u}_t\}&=\lim_{t\to \infty} \mathbb{E}\{ \hat{\bm{x}}_t^\intercal \mat{K}^\intercal \mat{R}_0\mat{K} \hat{\bm{x}}_t\}\\
        &=\trace{\mat{S}\mat{B}^*\mat{R}^{-1}\mat{R}_0\mat{R}^{-1}\mat{B}\mat{S}\mat{F}}.\qedhere
    \end{aligned}
\end{equation*}
\end{proof}
\begin{lemma}[Lemma 3,~\cite{de2004note}]
\label{lemma:invposef}
Let~$\mat{A},\mat{B}\in \posdef$,~$\mat{B}-\mat{A}\in \possemidef$. Then,~$\mat{A}^{-1},\mat{B}^{-1}\in \posdef$ and~$\mat{A}^{-1}-\mat{B}^{-1}\in \possemidef$.
\end{lemma}
\begin{comment}
\begin{lemma}
\label{lemma:codesigncnt_1}
Let~$\mat{Q}(\alpha)=\alpha \mat{Q}_0$ and~$\mat{R}(\alpha)=\frac{1}{\alpha}\mat{R}_0$, $\alpha\in \mathbb{R}_+$. Let~$\bm{u}^\star(\alpha)$ be the solution of the LQG problem with $\mat{Q}(\alpha)$ and $\mat{R}(\alpha)$. Then, under optimal control one has:
\begin{itemize}
    \item $P_\mathrm{track}(\alpha)$ is decreasing with $\alpha$
    increasing.
    \item $P_\mathrm{effort}(\alpha)$ is increasing with $\alpha$ increasing.
\end{itemize}
\end{lemma}
\end{comment}
\begin{proof}[Proof of \cref{lemma:codesigncnt_1}]
First:~$\alpha\leq \alpha'$,~$\mat{Q}=\alpha\mat{Q}_0\preceq \alpha'\mat{Q}_0=\mat{Q}'$ and~$\mat{R}=\mat{R}_0/\alpha\succeq \mat{R}'=\mat{R}_0/\alpha'$. For the first part, following \cref{lemma:precision}, we need to prove that
\begin{equation*}
\label{eq:first_part}
    \alpha\leq \alpha' \Rightarrow \trace{\mat{Q}_0\left( \mat{\Sigma} +\mat{F}\right)}\mid_{\alpha}\geq \trace{\mat{Q}_0\left( \mat{\Sigma} +\mat{F}\right)}\mid_{\alpha'}.
\end{equation*}
$\mat{\Sigma}$ is independent from~$\alpha$. From \cref{eq:lyapct} it is clear that~$\alpha\leq \alpha' \Rightarrow \mat{F}_1\succeq \mat{F}_2$, which confirms the statement. For the second part, following \cref{lemma:precision}, it is easy to prove that $\alpha\leq \alpha'$ implies
$\trace{ \mat{S}\mat{B}\alpha^2\mat{R}_0^{-1}\mat{B}^*\mat{S}\mat{F}}\mid_{\alpha} \leq \trace{ \mat{S}\mat{B}\alpha'^2\mat{R}_0^{-1}\mat{B}^*\mat{S}\mat{F}}\mid_{\alpha'}$.\qedhere
\end{proof}
\begin{comment}
\begin{lemma}
\label{lemma:cont_V_W_appendix}
The solution of \cref{eq:cont_ric_2} is monotonic in $\mat{V}$ and $\mat{W}$:
\begin{equation*}
\tup{\mat{V},\mat{W}'}\preceq \tup{\mat{V}',\mat{W}'} \Rightarrow \mat{\Sigma}(\mat{V},\mat{W})\preceq \mat{\Sigma}(\mat{V}',\mat{W}'). 
\end{equation*}
\end{lemma}
\end{comment}
\begin{proof}[Proof of \cref{lemma:cont_V_W}]
First, we know that $\bar{\mat{\Sigma}}$ satisfies \cref{eq:cont_ric_2} :
\begin{equation*}
    \mat{A\Sigma} + \mat{\Sigma A}^*-\mat{\Sigma C}^*\mat{V}^{-1}\mat{C\Sigma}+\mat{W}\coloneqq \mat{M}(\mat{V},\mat{W}).
    \end{equation*}
By letting $\tup{\mat{V}_1,\mat{W}_1}\preceq \tup{\mat{V}_2,\mat{W}_2}$, and using the notation $\mat{M}_i=\mat{M}(\mat{V}_i,\mat{W}_i)$, $i\in \{1,2\}$, from Theorem 3 in~\cite{de2004note} we know that $\mat{M}_2-\mat{M}_1 \succeq \mat{0} \Rightarrow \bar{\mat{\Sigma}}(\mat{V}_1,\mat{W}_1)\preceq \bar{\mat{\Sigma}}(\mat{V}_2,\mat{W}_2)$. We have:
    \begin{equation*}
    \begin{aligned}
        \mat{M}_2-\mat{M}_1&=-\mat{\Sigma}\mat{C}^*\left( \mat{V}_2^{-1}-\mat{V}_1^{-1}\right)\mat{C}\mat{\Sigma}+\mat{W}_2-\mat{W}_1\succeq \mat{0},
    \end{aligned}
    \end{equation*}
    where we use that $\mat{V}_1^{-1}-\mat{V}_2^{-1}\in \posdef$ and $\mat{W}_2-\mat{W}_1\in \posdef$ (\cref{lemma:invposef}).
    Therefore, $\bar{\mat{\Sigma}}$ is monotone in $\mat{V}$ and $\mat{W}$.
\end{proof}
\begin{comment}
\begin{lemma}
\label{lemma:codesigncnt_2}
Consider the situation of \cref{lemma:codesigncnt_1}:
\begin{itemize}
    \item Fix $P_\mathrm{track}$. $P_\mathrm{effort}$ is monotonic in $\mat{W}$ and in $\mat{V}$.
    \item Fix $P_\mathrm{effort}$. $P_\mathrm{track}$ is monotonic in $\mat{W}$ and in $\mat{V}$.
\end{itemize}
\end{lemma}
\end{comment}
\begin{proof}[Proof of \cref{lemma:codesigncnt_2}]
We need to prove:
\begin{equation*}
    \begin{aligned}
    \tup{\mat{V},\mat{W}}&\preceq \tup{\mat{V}',\mat{W}'}\Rightarrow \effort(\mat{V},\mat{W})\leq \effort(\mat{V}',\mat{W}'),\\
    \tup{\mat{V},\mat{W}}&\preceq \tup{\mat{V}',\mat{W}'}\Rightarrow \track(\mat{V},\mat{W})\leq \track(\mat{V}',\mat{W}').
    \end{aligned}
\end{equation*}
From \cref{lemma:cont_V_W}, we know $\mat{\Sigma}(\mat{V},\mat{W})\preceq \mat{\Sigma}(\mat{V}',\mat{W}')$. Since~$\mat{F}$ solves \cref{eq:lyapct}, we know $\mat{F}(\mat{V},\mat{W})\preceq \mat{F}(\mat{V}',\mat{W}')$. As no other term of $\effort$ or $\track$ depends on $\mat{V},\mat{W}$, we are done.
\end{proof}

\subsection{Proofs for \cref{sec:part_ii}}
\label{sec:app_B}
\begin{proof}[Proof of \cref{lemma:discretization}]
We consider the continuous-time dynamics given in \cref{def:ctlqg}, and sample this process with sampling period $\delta$. The input $\bm{u}_t$ is constant over the sampling period. We can write the solution of the dynamics as
\begin{equation}
\label{eq:stateequation}
    \bm{x}_t=\Phi(t,k\delta)\bm{x}_{k\delta}+\Gamma(t,k\delta)\bm{u}_{k\delta},
\end{equation}
where $\Phi(t,k\delta)$ satisfies
\begin{equation}
\label{eq:phidisc}
\frac{\D}{\D t} \Phi(t,k\delta)=\mat{A}\Phi(t,k\delta), \quad \Phi(k\delta,k\delta)=\mat{I},
\end{equation}
and
\begin{equation}
\label{eq:gammadisc}
    \Gamma(t,k\delta)=\smallint_{k\delta}^t\Phi(t,s)\bm{B}\D s.
\end{equation}
The sampled version of the dynamics is
\begin{equation*}
    \begin{aligned}
        \bm{x}_{k\delta+\delta}&=\Phi(t,k\delta)\bm{x}_{k\delta}+\Gamma(t,k\delta)\bm{u}_{k\delta}+\bm{w}_{k\delta}\\
        \bm{y}_{k\delta}&=\bm{C}\bm{x}_{k\delta}+\bm{v}_{k\delta},
    \end{aligned}
\end{equation*}
with covariances $\dis{\mat{W}}=\int_{0}^\delta e^{\mat{A}\tau}\mat{W}e^{\mat{A}^\intercal \tau}\D \tau$, and $\dis{\mat{V}}$. We can now manipulate the continuous-time cost provided in \cref{eq:general_continuous_cost}:
\begin{equation}
\label{eq:costtransformed}
    \begin{aligned}
        J&=\textstyle{\lim_{T\to \infty}} \frac{1}{T}\mathbb{E}\{\smallint_{0}^{T} \left(\left( \bm{x}_t^\intercal \mat{Q} \bm{x}_t\right)+\left( \bm{u}_t^\intercal \mat{R}\bm{u}_t\right)\right) \D t\}\\
        &=\textstyle{\lim_{N\to \infty}} \frac{1}{N}\mathbb{E}\{\sum_{k=0}^{N-1} \smallint_{k\delta}^{k\delta +\delta}\left(\left( \bm{x}_t^\intercal \mat{Q} \bm{x}_t\right)+\left( \bm{u}_t^\intercal \mat{R}\bm{u}_t\right)\right) \D t  \}.
    \end{aligned}
\end{equation}
We can now use \cref{eq:stateequation} to get:
\begin{equation*}
    \begin{aligned}
        &\smallint_{k\delta}^{k\delta + \delta}\left( \left( \bm{x}_t^\intercal \mat{Q} \bm{x}_t\right)+\left( \bm{u}_t^\intercal \mat{R}\bm{u}_t\right)\right) \D t= \bm{x}_{k\delta}^\intercal \dis{\mat{Q}}\bm{x}_{k\delta}+\bm{u}_{k\delta}^\intercal \dis{\mat{R}}\bm{u}_{k\delta},
    \end{aligned}
\end{equation*}
\begin{equation*}
    \begin{aligned}
    \dis{\mat{Q}}&=\smallint_{k\delta}^{k\delta + \delta} \Phi^\intercal(s,k\delta) \mat{Q}\Phi(s,k\delta) \D s,\\
    \dis{\mat{R}}&=\smallint_{k\delta}^{k\delta + \delta}\left( \Gamma^\intercal(s,k\delta) \mat{Q}\Gamma(s,k\delta)+\mat{R}\right)\D s.
    \end{aligned}
\end{equation*}
Hence,~\cref{eq:costtransformed} is $\underset{N\to \infty}{\lim} \frac{1}{N}\mathbb{E} \{ \sum_{k=0}^{N-1}\bm{x}_{k\delta}^\intercal \dis{\mat{Q}}\bm{x}_{k\delta}+\bm{u}_{k\delta}^\intercal \dis{\mat{R}}\bm{u}_{k\delta}\}$,
which indeed corresponds to the cost of a discrete-time LQG (\cref{eq:general_discrete_cost}). Solving \cref{eq:phidisc} one finds $\dis{\mat{A}}=\Phi(t,k\delta)=e^{\mat{A}\delta}$,
and one can hence write \cref{eq:gammadisc} as $\dis{\mat{B}}=\Gamma(\delta,0)=\smallint_{0}^\delta e^{\mat{A}s} \mat{B} \D s$.
\end{proof}

\subsubsection*{Proof of \cref{thm:codesigndt}}
\begin{lemma}
The optimal control law for a digital LQG problem is~$\bm{u}_k^\star =-\dis{\mat{K}}\hat{\bm{x}}_k=-\left( \dis{\mat{B}}^*\bar{\mat{P}}\dis{\mat{B}}+\dis{\mat{R}}\right)^{-1}\dis{\mat{B}}^*\bar{\mat{P}}\dis{\mat{A}} \hat{\bm{x}}_k$,
where $\bar{\mat{P}}\in \posdef$ is the solution of the Riccati equation
\begin{equation}
\label{eq:disc_ric_2}
    \mat{P}=\dis{\mat{A}}^* \mat{P}\dis{\mat{A}}-(\dis{\mat{A}}^* \mat{P}\dis{\mat{B}})(\dis{\mat{R}}+\dis{\mat{B}}^* \mat{P} \dis{\mat{B}})^{-1}(\dis{\mat{B}}^* \mat{P}\dis{\mat{A}})+\dis{\mat{Q}}.
\end{equation}
The minimum cost $\dis{J}^\star$ achieved by the optimal control is\footnote{Note that also~\cite{hendricks2008} contains a typo at p. 476 (- instead of + in Eq. 7.199).}:
\begin{equation*}
\label{eq:discmincost}
\begin{aligned}
&\dis{J}^\star=\trace{\left(\bar{\mat{P}}+\dis{\mat{Q}}\right)\dis{\mat{L}}\left( \dis{\mat{C}}\bar{\mat{\Gamma}}\dis{\mat{C}}^*+\dis{\mat{V}}\right)\dis{\mat{L}}^*+\dis{\mat{Q}}\bar{\mat{\Gamma}}}\\
&=\mathsf{Tr}\left((\dis{\mat{Q}}+\bar{\mat{P}})\dis{\mat{W}}+\bar{\mat{\Gamma}}\dis{\mat{K}}^*\left( \dis{\mat{R}}+\dis{\mat{B}}^*(\dis{\mat{Q}}+\bar{\mat{P}})\dis{\mat{B}}\right)\dis{\mat{K}}\right),
\end{aligned}
\end{equation*}
where $\bar{\mat{\Gamma}}\in \posdef$ is the solution of the Riccati equation
\begin{equation}
\label{eq:disc_ric_1}
    \mat{\Gamma}=\dis{\mat{A}}\mat{\Gamma}\dis{\mat{A}}^*-\dis{\mat{A}}\mat{\Gamma}\mat{C}^*(\mat{C}\mat{\Gamma}\mat{C}^*+\dis{\mat{V}})^{-1}\mat{C}\mat{\Gamma}\dis{\mat{A}}^*+\dis{\mat{W}}.
\end{equation}
\end{lemma}

\begin{lemma}
\label{lemma:tracking_dt}
One can write
\begin{equation*}
\begin{aligned}
    \textstyle{\lim_{k \to \infty}}\mathbb{E}\{\bm{x}_k^\intercal  \mat{Q}\bm{x}_k\}&=\trace{\mat{Q}(\mat{\Gamma}+\mat{F})},\\
    \textstyle{\lim_{k\to \infty}}\mathbb{E}\{ \bm{u}_k^\intercal \mat{R}\bm{u}_k\}&=\trace{\dis{\mat{K}}^\intercal \mat{R} \dis{\mat{K}}\mat{F}},
\end{aligned}
\end{equation*}
where $\mat{\Gamma}$ solves \cref{eq:disc_ric_1} and $\mat{F}$ solves the Lyapunov equation
\begin{equation}
\label{eq:lyapdis}
    \mat{F}=\dis{\mat{L}}\left( \mat{C}\mat{\Gamma}\mat{C}^* +\dis{\mat{V}}\right)\dis{\mat{L}}^\intercal +(\dis{\mat{A}}-\dis{\mat{B}}\dis{\mat{K}})\mat{F} (\dis{\mat{A}}-\dis{\mat{B}}\dis{\mat{K}})^*,
\end{equation}
with $\dis{\mat{L}}=\mat{\Gamma}\mat{C}^*(\mat{C}\mat{\Gamma}\mat{C}^*+\dis{\mat{V}})^{-1}$ discrete Kalman gain.
\end{lemma}

\begin{lemma}
\label{lemma:codesigndis_1}
Consider the LQG problem of \cref{lemma:discretization}. Let $\mat{Q}(\alpha)=\alpha \mat{Q}_0$ and $\mat{R}(\alpha)=\frac{1}{\alpha}\mat{R}_0$, $\alpha\in \mathbb{R}_+$. Let $\bm{u}^\star(\alpha)$ be the solution of the LQG problem with $\mat{Q}(\alpha)$ and $\mat{R}(\alpha)$. Then, under optimal control one has:
\begin{itemize}
    \item $\track^{\mathrm{d}}(\alpha)$ is decreasing with $\alpha$
    increasing.
    \item $\effort^\mathrm{d}(\alpha)$ is increasing with $\alpha$ increasing.
\end{itemize}
\end{lemma}
\begin{proof}
    The proof is analogous to the one of \cref{lemma:codesigncnt_1}.
\end{proof}

\begin{lemma}
\label{lemma:codesigndis_2}
Consider the situation of \cref{lemma:codesigndis_1}. One has:
\begin{itemize}
    \item Fix $\track^\mathrm{d}$. $\effort^\mathrm{d}$ is monotonic in $\dis{\mat{W}}$ and in $\dis{\mat{V}}$.
    \item Fix $\effort^\mathrm{d}$, $P_\mathrm{track}^\mathrm{d}$ is monotonic in $\dis{\mat{W}}$ and in $\dis{\mat{V}}$.
\end{itemize}
\end{lemma}
\begin{proof}
Consider $\tup{\dis{\mat{V}},\dis{\mat{W}}}\preceq \tup{\dis{\mat{V}}',\dis{\mat{W}}'}$.
From the \emph{Comparison Theorem} in~\cite{freiling2003}, we know that the solution $\bar{\mat{\Gamma}}$ of \cref{eq:disc_ric_1} is monotonic in both $\dis{\mat{V}}$ and $\dis{\mat{W}}$. Furthermore, the solution of \cref{eq:disc_ric_2} does not depend on $\dis{\mat{V}},\dis{\mat{W}}$. Finally, $\dis{\mat{F}}$ solves the discrete Lyapunov equation \cref{eq:lyapdis} and analogously to \cref{lemma:codesigncnt_2}, one can prove~$\mat{F}(\dis{\mat{V}},\dis{\mat{W}})\preceq \mat{F}(\dis{\mat{V}}',\dis{\mat{W}}')$.
\end{proof}

\begin{comment}
\input{ACC/tikz/dt_dt_cnt.tikz}

\input{ACC/tikz/ct_ct_cnt.tikz}

\input{ACC/tikz/ct_ct_delay_cnt.tikz}
\end{comment}

\begin{comment}
\newpage

\scalebox{0.7}{\begin{tikzpicture}
\node[block] (cnt) at (0,0) {Control};
\node[block, right=2cm of cnt] (act) {Steering};
\node[block, right=2cm of act] (plant) {AV};
\node[block, below=1cm of cnt] (est) {Estimation};
\node[block, below=1cm of plant] (sen) {Cameras};
\draw[->] (cnt) -- (act) node[pos=0.4,above]{command};
\draw[->] (act) -- (plant) node[pos=0.5,above]{control input};
\draw[->] (sen) -- (est) node[pos=0.5,below]{measurements};
\draw[->] (est) -- (cnt) node[pos=0.5,right]{state estimate};
\draw[->] ($(sen.east)+(0.75,-0.2)$) -- ($(sen.east)+(0,-0.2)$) node[pos=0,right]{\begin{tabular}{c}observation\\ noise\end{tabular}};
\draw[->] ($(plant.east)+(0,0)$) -- ($(plant.east)+(0.75,0)$) node[pos=1,right]{output};
\draw[->] ($(plant.east)+(0.4,0)$) -- ($(plant.east)+(0.4,-1.5)$)|- ($(sen.east)+(0,0.2)$);
\draw[->] ($(act.east)+(0.25,0)$) -- ($(act.east)+(0.25,-1.5)$)|- ($(est.east)+(0,0.2)$);
\draw[->] ($(plant.north)+(0,0.5)$) -- ($(plant.north)+(0,0)$)node[pos=0.5,right]{system noise};
\draw[dotted,blue,thick] (-1,-2.5) rectangle (2.25,0.5) {};
\node[blue] at (0.625,-2.3) {Computing};

\end{tikzpicture}}
\end{comment}

\bibliographystyle{IEEEtran}
\bibliography{paper}
}
{

\section{Conclusions}
\label{sec:conclusion}
We considered the co-design of robotic platforms and linear control systems steering them. We have shown how to frame LQG control problems in a monotone theory of co-design, framing them as monotone feasibility relations (\glspl{abk:dpi}). The approach guarantees modularity and compositionality, allowing us to model complex, interconnected systems and to capture heterogeneous design abstraction levels, ranging from control synthesis to hardware selection.
\paragraph*{Outlook}
This work fits in our vision of compositional robotics, where we want to describe and co-design robots all the way from signals characterising a single component~\cite{zardini2020compositional} to networks and fleet effects~\cite{ZardiniEtAlBis2020}. We want to leverage compositionality properties of our framework to interconnect the co-design problem of a robot in the co-design problem of an urban robot-enabled mobility system, to assess how local trade-offs influence fleet-level operations~\cite{choudhury2020efficient}. Furthermore, we want to extend the proposed studies to the synthesis of more complex control structures.

\section*{Acknowledgments}
The authors would like to thank Dr. Sawyer Fuller, Nicolas Lanzetti, Dr. Riccardo Bonalli, and Dr. Takashi Tanaka for the fruitful discussions.
\bibliographystyle{IEEEtran}
\bibliography{paper}}
\end{document}